\documentclass[aps,prd,showpacs,tighten,nofootinbib]{revtex4}
\usepackage{amssymb}

\usepackage{graphicx}
\usepackage{bm}
\usepackage{amsmath}
\usepackage{latexsym}

\input{tcilatex}

\begin{document}

\title{Vacuum polarization of  massive spinor and vector fields in the
spacetime of a nonlinear black hole }

\author{Jerzy Matyjasek}
\email{matyjase@tytan.umcs.lublin.pl, jurek@kft.umcs.lublin.pl}

\affiliation{Institute of Physics, 
Maria Curie-Sk\l odowska University\\
pl. Marii Curie-Sk\l odowskiej 1, 
20-031 Lublin, Poland}

\begin{abstract}
Building on general formulas obtained from the approximate  renormalized 
effective action, the  stress-energy tensor of the quantized massive 
spinor and vector fields in the spacetime of the regular black hole 
is constructed. Such a black hole is the solution to the coupled
system of nonlinear electrodynamics and general relativity. A detailed 
analytical and numerical analysis of the stress-energy tensor in the exterior 
region is presented. It is shown that for small values of the charge as 
well as large distances from the black hole the leading behavior 
of the stress-energy tensor is similar to that in the Reissner-Nordstr\"om
geometry. Important differences appear when the inner horizon becomes 
close to the event horizon. A special emphasis is put on the extremal 
configuration and it is shown that the stress-energy tensor is regular 
inside the event horizon of the extremal black hole.
\end{abstract}


\maketitle









\section{\label{intro}Introduction}

One the most important and intriguing open questions in the physics of
compact objects is the issue of the final stage of black hole
evolution and the problem of singularities residing in the internal
region of the black holes. Although the definitive answer to these 
questions would require the application of the full machinery 
of the (unknown as yet) quantum gravity or
even more sophisticated approach, one can still obtain 
valuable results within the semiclassical framework. Of course, the
equations of the semiclassical gravity cannot be used to describe the
evolution of the system completely: such equations are expected to break
down in the Planck regime. On the other hand, however, having established
the domain of applicability of the theory precisely one can obtain 
interesting and important results. Moreover, a careful analysis of the solutions 
to the
semiclassical equations can show us tendencies in the evolution of the
system, indicating its possible continuation.

It is well known that the physical content of quantum field theory in
spacetimes describing black holes is carried by the renormalized
stress-energy tensor (SET) evaluated in a suitable state~\cite{Birrell}. Treating
the renormalized stress-energy tensor as the source term of the
semiclassical Einstein field equations, one can, in principle, determine the
back reaction of the quantized fields upon the spacetime geometry of black
holes unless the (expected) quantum gravity effects become dominant.
Therefore, form the point of view of the semiclassical approach, it is
crucial to have at one's disposal a general formula describing functional
dependence of the renormalized stress-energy tensor on a wide class of
metrics. 

In the semi-classical approach we are confronted with
two major problems: construction of the renormalized stress-energy tensor on the
one hand, and studying its influence via semi-classical equations on the
system on the other. Unfortunately, even such a simplified approach leads to
the equations that are  still far too complicated to be solved exactly
and it is natural that much effort has been concentrated on developing
approximate methods, referring to numerical calculations or both.

It seems that for the massive fields, the approximation based on the
Schwinger-DeWitt expansion~\cite{Schwinger,DeWitt,Vilkovisky} is of required
generality. Indeed, it has been shown that for sufficiently massive fields,
the renormalized effective action, $W_{ren}^{\left( 1\right) }$, can be
expanded in powers of $m^{-2}.$ It is because the nonlocal contribution to
the total effective action can be neglected, and, consequently, there remains only 
the  vacuum polarization part which is local and determined by 
the geometry of the spacetime.
The stress-energy tensor can, therefore, be obtained by functional differentiation of
the effective action with respect to the metric tensor: 
\begin{equation}
{\frac{2}{g^{1/2}}}{\frac{\delta }{\delta g_{\mu \nu }}}W_{ren}^{(1)}\,=\,%
\langle T^{\mu \nu }\rangle _{ren}.  \label{stress}
\end{equation}
Such a tensor describing the vacuum polarization effects of
the quantized massive fields in the vacuum type-D geometries  
has been constructed and subsequently applied in a series of
papers by Frolov and Zel'nikov~\cite{fz82,fz83,fz84massiv}. 
They used the Schwinger - DeWitt
method~\cite{Schwinger,DeWitt} and constructed the first order approximation of
the effective action, omitting the terms which do not contribute to the final 
results in the
Ricci-flat spaces. These results have subsequently been extended
in Refs. \cite{Jirinek00prd} and \cite{Jirinek01prd1},
where the most general formulas describing the renormalized stress-energy
tensor of the massive scalar, spinor and vector fields have been calculated.
As  the effective action consists of 10 (integrated) purely geometric
terms constructed from the curvature tensor and multiplied   by the spin-dependent
coefficients, it suffices to calculate their functional derivatives
with respect to metric tensor only once. The stress-energy tensor of 
the scalar, spinor and vector  fields can easily be obtained by 
taking the linear combination of the thus obtained functional derivatives
with the spin-dependent coefficients. Interested reader is referred 
to~\cite{Jirinek00prd} and~\cite{Jirinek01prd1}. (Especially see Eqs. 7-18 and Table I 
of  Ref~\cite{Jirinek01prd1}).

The range
of applicability of such a stress-energy tensor is dictated by
the limitations of the validity of the renormalized effective action: 
it can be
used in any spacetime provided the mass of the quantized field is
sufficiently great, i.e., when the Compton length, $\lambda_{c},$ 
is much smaller than the characteristic radius of curvature, $L,$ 
where the latter means any
characteristic length scale associated with the geometry in question.
Assuming, for example, that $L$ is related to the Kretschmann scalar as
\begin{equation}
K =  R_{\mu \nu \sigma \tau}R^{\mu \nu \sigma \tau} \sim L^{-4},
\end{equation}
one has a simple criterion for the validity of the Schwinger-DeWitt
expansion. Typically, $L \sim M_{H},$
where $M_{H}$ is the black hole mass, and, therefore, one expects 
that the approximation would 
be accurate provided $m$ and $L$ satisfy $m L \sim M_{H}m \gg 1.$

Using a different method, Anderson, Hiscock, and Samuel \cite
{Anderson1,Anderson2} evaluated $\langle T_{\nu }^{\mu }\rangle _{ren}$ of
the massive scalar field with arbitrary curvature coupling for a general
static, spherically symmetric spacetime and applied the obtained formulas to the
Reissner-Nordstr\"{o}m (RN) spacetime. (See also Ref.~\cite{Popov}.)
Their approximation is equivalent to the
Schwinger - DeWitt expansion; to obtain the lowest (i. e. $m^{-2}$) terms,
one has to use sixth-order WKB expansion of the mode functions. Numerical
calculations reported in Ref. \cite{Anderson1,Anderson2} indicate that the
Schwinger-DeWitt method always provide a good approximation of the
renormalized stress energy tensor of the massive scalar field with arbitrary
curvature coupling as long as the mass of the field remains sufficiently
large. The techniques presented in refs.~\cite{Anderson1,Anderson2} have been 
successfully applied in a number of cases. 
Specifically, the vacuum polarization in the electrically charged
black holes have been studied in Refs.~\cite{Anderson1,Anderson2},
important issue of the black hole interiors in \cite{Anderson_interior}, the
stress-energy tensor in the spacetimes of various wormhole types in~\cite
{Anderson_wormhole} and the back reaction calculations in~\cite{Anderson4}.

On the other hand, the general formulas of Refs.~\cite{Jirinek00prd,Jirinek01prd1} 
have been applied in the spactimes of the Reissner-Nordstr\"om~\cite{Jirinek00prd} 
and dilaton black holes~\cite{Matyjasek:2002rp}.  
Various aspects of the back reaction problem have been studied in 
Refs.~\cite{extreme,jirinek03b,Matyjasek:2005+}.
Especially interesting in this regard are the regular black hole
geometries, being the solutions of the coupled system of equations of the
nonlinear electrodynamics and gravitation. The stress-energy tensor of the
massive scalar fields (with an arbitrary curvature coupling) 
in such spaces have been studied in~\cite{Jirinek01prd1,Jirinek02prd}.

The issue of the regular black holes in general relativity has a long and
interesting history. For example, one of the methods that can be used in
construction of such configurations consits in replacing the singular black
hole interior by a regular core. This idea appeared
in mid sixties~\cite{Sakharov,Gliner,Bardeen:68} and its various realizations 
have been and still are investigated. For example, in the models considered in 
Refs.~\cite{frolov1,frolov2} part of the region inside the event horizon 
is joined through a thin boundary layer to de Sitter geometry. Similar
idea has been applied in the calculations reported in Ref.~\cite{Oleg_BB},
where the singular interior of the extreme Reissner-Nordstr\"{o}m black hole 
has been replaced by the Bertotti-Robinson geometry.
Of course, such a geometric surgery does not exhaust all interesting possibilities. 
The regular geometries constructed with the aid of  suitably chosen 
profile functions, or, better, the exact 
solutions constructed for specific, physically reasonable sources are 
of equal importance~\cite{Irina1,Irina2,Borde,Mars,Ayon-Beato:2000zs,
Ayon-Beato:2004ih}.

Recent interest in the nonlinear electrodynamics is partially motivated
(beside a natural curiosity) by the fact that the theories of this type
frequently arise in modern theoretical physics. For example, they appear as
effective theories of string/M-theory. 
Moreover, on general grounds one expects that it should
be possible to construct the regular black hole solutions 
to the coupled system of equations of the nonlinear electrodynamics and gravity. 

One of the most interesting and intriguing regular solutions has been constructed by
Ay\'{o}n-Beato and Garc\'{i}a \cite{ABG}.  It has subsequently been 
reinterpreted by Bronnikov in Refs.~\cite{Bronnikov1,Bronnikov2}.
The former solution describes a regular static and spherically symmetric 
configuration parametrized by the  mass and the
electric charge whereas the latter describes a formally similar geometry
characterized by $M$ and the magnetic charge $Q$. 
We shall refer to the solutions of this kind as ABGB geometries.
It should be noted that the electric solution does not contradict the no-go
theorem, which states that if the Lagrangian of the matter fields, ${\cal L},$ 
is an arbitrary function
of $F=F_{\mu\nu}F^{\mu\nu}$ with the Maxwell asymptotics in a weak field limit 
($F_{\mu\nu}$ is the electromagnetic tensor), then 
it cannot have a regular center.
This is because the formulation of the nonlinear electrodynamics employed by
Ay\'{o}n-Beato and Garc\'{i}a ($\mathcal{P}$ framework in the nomenclature
of Refs. \cite{Bronnikov1,Bronnikov2}) is not the one to which one refers in
the assumptions of the theorem. Specifically, the solution of Ay\'{o}n-Beato
and Garc\'{i}a has been constructed in a formulation of the nonlinear
electrodynamics obtained from the original one ($\mathcal{F}$ framework) by
means of a Legendre transformation (see Ref.~ \cite{Bronnikov2} for
details). 

For certain values of the parameters the  ABGB line element  
describes a black hole and an attractive feature of this solutions 
that simplifies calculations is the possibility to express 
the location of the horizons in terms
of the Lambert special functions \cite{Jirinek01prd1,Jirinek02prd}.
Moreover, as the function ${\cal L}(F)$  coincides with the Lagrangian density 
of the Maxwell theory in the weak field limit, one expects that at large distances 
the static and spherically symmetric solution should approach the
Reissner-Nordstr\"{o}m solution. A similar behavior should occur outside the
event horizon for $\left| e\right| /M\ll 1,$ where $e$ denotes either the electric or magnetic 
charge and $M$ is the mass.

The objective of this paper is to construct the renormalized stress-energy
tensor of the quantized neutral massive spinor and vector fields in the
spacetime of the regular ABGB black hole. The stress-energy tensor of the
scalar fields with arbitrary curvature coupling has been constructed and
discussed in Refs.~\cite{Jirinek01prd1,Jirinek02prd}. The results presented
here are the basic ingredients of the first-order back reaction calculations.
They can also be used in the analysis of the various energy conditions and
quantum inequalities.

\section{The renormalized effective action}
The source term of the semi-classical Einstein field equations is given by the
stress-energy tensor. Ideally, such a tensor should be constructed from 
the renormalized effective action, $W_{eff},$ in a standard way, i.e., by
functional differentiation of $W_{eff},$ with respect to the metric.
Unfortunately, neither the exact nor the approximate form of $W_{eff}$ 
is known in general. However, in a large mass limit of the quantized fields
one can construct its local approximation satisfactorily describing the vacuum
polarization effects.

The massive scalar, spinor and vector fields in curved spacetime satisfy
the differential equations
\begin{equation}
(-\,\Box \,+\,\xi R\,+\,m^{2})\phi ^{(0)}\,=0,  \label{s0}
\end{equation}
\begin{equation}
(\gamma ^{\mu }\nabla _{\mu }\,+\,m)\phi ^{(1/2)}\,=\,0  \label{s12}
\end{equation}
and 
\begin{equation}
(\delta _{\nu }^{\mu }\Box \,-\,\nabla _{\nu }\nabla ^{\mu }\,-\,R_{\nu
}^{\mu }\,-\,\delta _{\nu }^{\mu }m^{2})\phi ^{(1)}\,=\,0,  \label{s1}
\end{equation}
respectively, 
where $\xi $ is the curvature coupling constant, and $\gamma ^{\mu }$ are the Dirac
matrices obeying standard relations $\gamma ^{\alpha }\gamma ^{\beta
}\,+\,\gamma ^{\beta }\gamma ^{\alpha }\,=\,2\hat{1}g^{\alpha \beta }.$ 
The lowest-order approximation of the renormalized effective action, $W_{ren}^{(1)},$ of 
the quantized massive fields satisfying equations (\ref{s0}-\ref{s1})
is given by a remarkably simple expression
\begin{equation}
W_{ren}^{(1)}\,=\,{\frac{1}{32\pi ^{2}m^{2}}}\int g^{1/2}d^{4}x\left\{ 
\begin{array}{l}
\lbrack a_{3}^{(0)}] \\ 
-tr[a_{3}^{(1/2)}] \\ 
tr[a_{3}^{(1)}]\,-\,[a_{3|\xi =0}^{(0)}]
\end{array}
\right.   \label{Weff}
\end{equation}
Here $[a_{3}^{\left( s\right) }]$ is the coincidence limit of the fourth
Hadamard-DeWitt-Minakshisundaram-Seeley~\cite{Gilkey} coefficient of the
scalars ($s=0),$ spinors ($s=1/2$) and vectors ($s=1)$. Making use of
elementary properties of the Dirac matrices and the Riemann tensor, after simple
calculations, one obtains the first term of the asymptotic expansion of the
effective action in the form \cite{Avra1,Avra3} 
\begin{eqnarray}
W_{ren}^{(1)}\, 
&=& {\frac{1}{192\pi ^{2}m^{2}}} \sum_{i=1}^{10} \alpha^{(s)}_{i} W_{i}\nonumber\\
&=&\,{\frac{1}{192\pi ^{2}m^{2}}}\int d^{4}xg^{1/2}\left(
\alpha _{1}^{(s)}R\Box R\,+\,\alpha _{2}^{(s)}R_{\mu \nu }\Box R^{\mu \nu
}\,+\,\alpha _{3}^{(s)}R^{3}\,+\,\alpha _{4}^{(s)}RR_{\mu \nu }R^{\mu \nu
}\right.   \notag \\
&&+\,\alpha _{5}^{(s)}RR_{\mu \nu \rho \sigma }R^{\mu \nu \rho \sigma
}\,+\,\alpha _{6}^{(s)}R_{\nu }^{\mu }R_{\rho }^{\nu }R_{\mu }^{\rho
}\,+\,\alpha _{7}^{(s)}R^{\mu \nu }R_{\rho \sigma }R_{~\mu ~\nu }^{\rho
~\sigma }  \notag \\
&&+\left. \,\alpha _{8}^{(s)}R_{\mu \nu }R_{\lambda \rho \sigma }^{\mu
}R^{\nu \lambda \rho \sigma }\,+\,\alpha _{9}^{(s)}{R_{\rho \sigma }}^{\mu
\nu }{R_{\mu \nu }}^{\lambda \gamma }{R_{\lambda \gamma }}^{\rho \sigma
}\,+\,\alpha _{10}^{(s)}R_{~\mu ~\nu }^{\rho ~\sigma }R_{~\lambda ~\gamma
}^{\mu ~\nu }R_{~\rho ~\sigma }^{\lambda ~\gamma }\right) 
\label{eff_act_2}
\end{eqnarray}
where the numerical coefficients $\alpha _{i}^{\left( s\right) }$ depending
on the spin of the field are given in a Table I.

Up to now, we have not specified the quantum state of the field.
However, the construction of the effective action has been carried out with the
assumption that the state in question may be identified with the Hartle-Hawking
state. A closer examination of the problem indicates that outside the narrow strip in
the closest vicinity of the event horizon, the results obtained in the Hartle-Hawking as
well as the Unruh and the Boulware states are almost indistinguishable as they differ by 
the contributions of the real particles. On the other hand, inside that region 
the stress-energy tensor strongly depends on the chosen state and may diverge at the
event horizon. On general grounds, one expects that for regular geometries
the Schwinger-DeWitt approximation yields a regular stress-energy tensor at 
the event horizon.

It should be stressed that although the effective action $W_{ren}^{(1)}$ 
can, in principle, be calculated for any line element, its physical 
applications are limited to the quantized fields in the large mass limit. 
Moreover, the technical difficulties one may encounter in the process of
calculation may prevent direct application of the effective action and 
the stress-energy tensor. Finally, observe that the effective action approach 
employed in this paper requires the metric of the spacetime to be positively
defined. Hence, to obtain the physical stress-energy tensor one has to analitically
continue the results constructed for the Euclidean metric.

\begin{table}[tbp]
\caption{The coefficients $\protect\alpha_{i}^{(s)}$ for the massive scalar,
spinor, and vector field}
\label{table1}
\begin{tabular}{cccc}
& s = 0 & s = 1/2 & s = 1 \\ 
$\alpha^{(s)}_{1} $ & ${\frac{1}{2}}\xi^{2}\,-\,{\frac{1}{5}} \xi $\,+\,${%
\frac{1}{56}}$ & $- {\frac{3}{280}}$ & $- {\frac{27}{280}}$ \\ 
$\alpha^{(s)}_{2}$ & ${\frac{1}{140}}$ & ${\frac{1}{28}}$ & ${\frac{9 }{28}}$
\\ 
$\alpha^{(s)}_{3}$ & $\left( {\frac{1}{6}} - \xi\right)^{3}$ & ${\frac{1}{864%
}}$ & $- {\frac{5}{72}}$ \\ 
$\alpha^{(s)}_{4}$ & $- {\frac{1}{30}}\left( {\frac{1}{6}} - \xi\right) $ & $%
- {\frac{1}{180}}$ & ${\frac{31}{60}}$ \\ 
$\alpha^{(s)}_{5}$ & ${\frac{1}{30}}\left( {\frac{1}{6}} - \xi\right)$ & $- {%
\frac{7}{1440}}$ & $- {\frac{1}{10}}$ \\ 
$\alpha^{(s)}_{6}$ & $- {\frac{8}{945}} $ & $- {\frac{25 }{756}}$ & $- {%
\frac{52}{63}}$ \\ 
$\alpha^{(s)}_{7}$ & ${\frac{2 }{315}}$ & ${\frac{47}{1260}}$ & $- {\frac{19}{%
105}} $ \\ 
$\alpha^{(s)}_{8}$ & ${\frac{1}{1260}}$ & ${\frac{19}{1260}} $ & ${\frac{61}{%
140}} $ \\ 
$\alpha^{(s)}_{9}$ & ${\frac{17}{7560}}$ & ${\frac{29}{7560}}$ & $- {\frac{67%
}{2520}}$ \\ 
$\alpha^{(s)}_{10}$ & $- {\frac{1}{270}}$ & $- {\frac{1}{108}} $ & ${\frac{1}{%
18}}$%
\end{tabular}
\end{table}

\section{The regular ABGB black hole}

An interesting solution to the coupled system of nonlinear
electrodynamics and gravity representing a class of the black holes 
parametrized by a mass and a charge has been constructed 
recently by Ay\'{o}n-Beato
and Garc\'{i}a~\cite{ABG} and by Bronnikov \cite{Bronnikov1,Bronnikov2}. The former
describes electrically charged configuration in the $\mathcal{P}$-framework
whereas the latter describes geometry of the magnetically charged solution
in the $\mathcal{F}$-framework. Both line elements are formally
identical and can be written in the form 
\begin{equation}
ds^{2}=-f\left( r\right) dt^{2}+f^{-1}\left( r\right) dr^{2}+r^{2}\left(
d\theta ^{2}+\sin ^{2}\theta d\phi ^{2}\right) ,  \label{line_abg}
\end{equation}
where 
\begin{equation}
f(r)\,=\,1-{\frac{2M_{H}}{r}}\left[ 1-\tanh \left( {\frac{e^{2}}{2M_{H} r}}\right) 
\right],  \label{gg}
\end{equation}
$M_{H}$ is the black hole mass and $e$ is either the magnetic  
or the electric charge.
For small values of the charge it differs outside the event horizon 
from the Reissner-Nordstr\"{o}m
solution by terms of order $\mathcal{O}(e^{6}).$ Similarly, at large distances the
the function $f(r)$ also closely resembles that of the RN
solution. Indeed, expanding metric potentials in a power series one
concludes that the ABGB solution behaves asymptotically as 
\begin{equation}
f(r)=\,1\,-\,{\frac{2M_{H}}{r}}\,+\,{\frac{e^{2}}{r^{2}}}\,-\,{\frac{e^{6}}{
12M_{H}^{2}r^{4}}}\,+\,\mathcal{O}({\frac{1}{r^{6}}}).  \label{exp}
\end{equation}

On the other hand, and this is even more interesting and has profound
consequences, the interior of the ABGB solution is regular. This can be
demonstrated by studying behavior of various curvature invariants.
It can be shown that curvature invariants factorize in such a way
that there is a common multiplicative factor,
which for $r \to 0$  behaves asymptotically  as $\exp\left( -e^2/M_{H} r\right).$
For $e=0$ the ABGB solution reduces to the Schwarzschild line element
and it is the nonlinear charge, no matter how small,
that leads to the dramatic changes of the geometry.

The spacetime described by the line element (\ref{line_abg}) with (\ref{gg})
has been extensively studied in~\cite{ABG,Jirinek01prd1,Jirinek02prd}.
Specifically, it has been shown that although the metric coefficient $f(r)$
is a complicated function of $r$, the location of the horizons may be
elegantly expressed in terms of the Lambert functions \cite{Jirinek01prd1}.
Since these results are, apparently, not widely known we shall 
summarize a few basic facts.
For a short description of the Lambert functions the reader is
referred to~\cite{Knuth1a}. 

Making use of the substitution $r=Mx$
and $e^{2}=q^{2}M^{2},$ and subsequently introducing a new unknown function $
W$ by means of the relation 
\begin{equation}
x=-{\frac{4q^{2}}{4W\,-\,q^{2}}},  \label{eq1}
\end{equation}
one arrives at 
\begin{equation}
\exp (W)W=-{\frac{q^{2}}{4}}\exp ({q^{2}/4}).  \label{eq2}
\end{equation}
Since the Lambert function is defined as 
\begin{equation}
\exp (W(s))W(s)\,=\,s,  \label{eq3}
\end{equation}
one concludes that the location of the horizons as a function of $q=|e|/M,$
is given by the real branches of the Lambert functions 
\begin{equation}
x_{+}\,=\,-{\frac{4q^{2}}{4W(0,-{\frac{q^{2}}{4}}\exp (q^{2}/4))-q^{2}}},
\label{eh}
\end{equation}
and 
\begin{equation}
x_{-}\,=\,-{\frac{4q^{2}}{4W(-1,-{\frac{q^{2}}{4}}\exp (q^{2}/4))-q^{2}}}.
\label{ih}
\end{equation}
The functions $W(0,s)$ and $W(-1,s)$ are the only real branches of the
Lambert function with the branch point at $s=-1/\mathrm{e},$ where e is the
base of natural logarithms. Finally, observe, that simple manipulations 
of Eqs. (\ref{eh}) and (\ref{ih}) indicate that for 
\begin{equation}
q_{extr}=\,2w^{1/2}=\,1.056,  \label{qextr}
\end{equation}
the horizons $r_{+}$ and $r_{-}$ merge at 
\begin{equation}
x_{extr}\,=\,{\frac{4w}{1+w}=}0.871,  \label{xextr}
\end{equation}
where $w=W\left( 1/\mathrm{e}\right) $ and $W(s)$ is a principal branch of
the Lambert function $W(0,s).$ 

Inspection of (\ref{qextr}) reveals another
interesting feature of the ABGB geometries: the black hole solution exists
for $q$ greater than the analogous ratio of the parameters of the RN
solution. 
The three types of the ABGB solutions therefore are: the regular black hole
with the inner and event horizons for $q<q_{extr},$ the extremal black hole
for $q=q_{extr},$ and the regular configuration for $q>q_{extr}.$ As have
been observed earlier at large distances as well as for small charges the
geometry of the ABGB solution resembles that of the Reissner-Nordstr\"{o}m.
There is, however, one notable distinction: for $q>1,$ the
Reissner-Nordstr\"{o}m solution describes unphysical naked singularity whereas the
regular geometry for $q>q_{extr}$ could be interpreted as a particle like
solution.

\section{Stress-energy tensor}

The renormalized stress-energy tensor of the quantized massive 
scalar (with an arbitrary curvature coupling), spinor and vector
fields in a large mass limit has a general form
\begin{equation}
\langle T^{\mu \nu}\rangle _{ren} = \frac{1}{96\pi^{2}m^{2}g^{1/2}}
\sum_{i=1}^{10} \alpha^{(s)}_{i}\frac{\delta}{\delta g_{\mu \nu}}W_{i},
\end{equation} 
where $W_{i}$ can be obtained from Eq.~(\ref{eff_act_2}) and the spin dependent 
coefficients are listed in Table I. The purely geometric objects 
$\delta W_{i}/\delta g_{\mu \nu}$ 
have been calculated in Refs.~\cite{Jirinek01prd1,Jirinek00prd}.
It has been shown that the thus obtained renormalized
stress-energy tensor consists 
of approximately 100 terms (constructed from the Riemann tensor, its covariant
derivatives and contractions) combined with the numerical coefficients depending on
the spin of the quantized field. However, such a local geometric structure 
of the stress-energy tensor has its price: $\langle T^{\mu \nu}\rangle _{ren}$
does not describe the process of particle creation which is a nonlocal
phenomenon. Fortunately, for sufficiently massive fields, the contribution
of the real particles can be neglected and the Schwinger-DeWitt action
satisfactorily approximates the total effective action. 

The general expression describing the renormalized stress-energy tensor of the
quantized fields in a large mass limit is rather complicated and to avoid
unnecessary proliferation of lengthy formulas it will be not presented here.
For its full form as well as the technical details the interested reader is
referred to \cite{Jirinek01prd1} (Especially see Eqs.7-18) and~\cite{Jirinek00prd}.
Because of numerous identities that hold for the Riemann tensor, the
final form of the stress-energy tensor is not unique and obviously
depends on adapted simplification strategies. 
It should be noted, however, that any other calculation based on the 
effective action (\ref{eff_act_2}) with the numerical 
coefficients $\alpha^{(s)}_{i}$ for scalar, spinor and vector field must 
yield results identical to those of Refs.~\cite{Jirinek00prd,Jirinek01prd1}. 
Recently, an equivalent form of the renormalized stress-energy tensor of the quantized
massive fields has been constructed by Folacci and Decanini~\cite{Folacci}. 

As has been stated earlier, the general formulas of Refs.~\cite{Jirinek00prd,Jirinek01prd1} 
have been successfully applied in a number cases, such as Reissner-Nordstr\"om
spacetime~\cite{Jirinek00prd}, dilatonic black holes~\cite{Matyjasek:2002rp},  
and various back reaction calculations~\cite{extreme,jirinek03b,Matyjasek:2005+}.
Moreover, the renormalized stress-energy tensor of the quantized massive 
scalar field with the arbitrary
curvature coupling in the ABGB geometry has been calculated and
exhaustively discussed in Refs.~\cite{Jirinek00prd,Jirinek01prd1}.
In this section we shall extend these calculations to the massive spinor and
vector fields. 

Since the general form of the stress-energy tensor is rather
complicated, one expects that its components evaluated for the specific line
element are, except simple geometries, formidable. 
Our calculations in the ABGB background clearly shows that this is indeed the case, and
once again, to avoid unnecessary proliferation of long formulas we shall not display them
here\footnote{The complete results in various formats can be obtained from the author}. 
On the other hand, one can obtain a great deal of information studying the
behaviour of the components of the stress-energy tensor in some physically important
regimes. Below we shall consider expansions of the
stress-energy tensor for small $q,$ large $x,$ and study the configuration near and at
the extremality limit. Special attention will be put on the regularity issues
and the interior of the extreme ABGB black hole.

\subsection{General features of the stress-energy tensor in the ABGB spacetime}

Each component of $\langle T_{\mu }^{\nu }\rangle _{ren}^{\left( s\right) }$
in the ABGB spacetime has a general form 
\begin{equation}
\langle T_{\mu }^{\nu }\rangle _{ren}^{\left( s\right) }=\frac{1}{96\pi
^{2}m^{2}M_{H}^{6}}\left( 1-\tanh \frac{q^{2}}{2x}\right) \sum_{i,j,k}\alpha
_{ijk}^{\left( s\right) }\frac{q^{2i}}{x^{j}}\tanh ^{k}\frac{q^{2}}{2x},
\label{structure}
\end{equation}
where $0\leq i\leq 6,$ $8\leq j\leq 15$, $0\leq k\leq 8,$ and $\alpha
_{ijk}^{\left( s\right) }$ are numerical coefficients depending on the spin
of the field. For simplicity, we have omitted tensor indices in right hand
side of the above equation. This result can be contrasted 
with the analogous expression
obtained for the Reissner-Nordstr\"{o}m black hole 
\begin{equation}
\langle T_{\mu }^{\nu }\rangle _{RN}^{\left( s\right) }=
\frac{1}{96\pi ^{2}m^{2}M_{H}^{6}}\sum_{i,j}\beta _{ij}^{\left( s\right) }
\frac{q^{2i}}{x^{j}},  \label{structure_rn}
\end{equation}
where $0\leq i\leq 3,$ $8\leq j\leq 12$ and $\beta _{ij}^{\left( s\right) }$
are, as before, the numerical coefficients depending on $s.$ In both cases
the stress-energy tensor is covariantly conserved and falls as $r^{-8
\text{ }}$as $r\rightarrow \infty .$ The latter behavior indicates that
there is no need to impose spherical boxes in the back reaction calculations.  
Moreover, both tensors are regular at the event horizon.

Since the Lagrangian density of the classical (nonlinear) field considered
in this paper tends to its Maxwell analogue as $F(=F_{\mu \nu }F^{\mu \nu
})\rightarrow 0,$ one expects that in this limit, regardless of the spin of
the quantized field, the leading behavior of the renormalized stress-energy
tensor of the quantized massive fields is similar to the analogous terms constructed
in the Reissner-Nordstr\"{o}m geometry. 
On the other hand, for the configurations near the extremality
limit the differences between the tensors outside the event horizon
should be more prominent.

It can be demonstrated that the difference between the radial and time
components of the stress-energy tensor factorizes: 
\begin{equation}
\langle T_{r}^{r}\rangle _{ren}^{\left( s\right) }-\langle T_{t}^{t}\rangle
_{ren}^{\left( s\right) }=\left[ 1-\frac{2}{x}\left( 1-\tanh \frac{q^{2}}{2x}%
\right) \right] F\left( x\right) ,  \label{factorize}
\end{equation}
where $F\left( x\right) $ is a regular function. Now, let us consider a
freely falling frame. A simple calculation shows that \ the frame components
of the tensor $T_{\mu }^{\nu }$ are 
\begin{equation}
T_{\left( 0\right) \left( 0\right) }=\frac{\gamma ^{2}\left(
T_{1}^{1}-T_{0}^{0}\right) }{f}-T_{1}^{1},
\end{equation}
\begin{equation}
T_{\left( 1\right) \left( 1\right) }=\frac{\gamma ^{2}\left(
T_{1}^{1}-T_{0}^{0}\right) }{f}+T_{1}^{1},
\end{equation}
\begin{equation}
T_{\left( 0\right) \left( 1\right) }=-\frac{\gamma \sqrt{\gamma ^{2}-f}
\left( T_{1}^{1}-T_{0}^{0}\right) }{f},
\end{equation}
where $\gamma $ is the energy per unit mass along the geodesic. One
concludes, therefore, that since all components of $\langle T_{\mu }^{\nu
}\rangle _{ren}^{\left( s\right) }$ are regular and \ $\left( \langle
T_{r}^{r}\rangle _{ren}^{\left( s\right) }-\langle T_{t}^{t}\rangle
_{ren}^{\left( s\right) }\right) /f$ is  by (\ref{factorize}) finite, the
stress-energy tensor of the quantize massive fields is regular in freely
falling frame.

\subsection{Stress-energy tensor on AdS$_{2}\times \mathrm{S}^{2}$ spacetime}

Let us postpone the detailed analysis of the stress-energy
tensor in the ABGB spacetime for a while and consider a far more simple case 
of the $AdS_{2}\times S^{2}$ geometry. Such geometries are closely related 
to the extremal black holes. Indeed, $AdS_{2}\times S^{2}$ can be obtained
by expanding the geometry 
of the vicinity of the event horizon into a whole manifold. Various
aspects of the geometries of this type have been discussed, for example, 
in~\cite{Bertotti,Robinson,
kofman1,Bardeen2,Oleg:1997,solconst,Jirinek00prd,Dias,Cardoso,
Dadhich,Caldarelli:2000wc,extreme,
Kocio_2004_n}.

The extremal ABGB black hole is described by a line element (\ref{line_abg})
with 
\begin{equation}
f(r)\,=\,1\,-\,\frac{2M_{H}}{r}\left[ 1\,-\,\tanh \left( \frac{2M_{H}w}{r}
\right) \right] .
\end{equation}
Now, in order to investigate the geometry in the vicinity of the event horizon, $
x_{extr}$ and to obtain uniform approximation we introduce new coordinates 
\begin{equation}
{
\tilde{t}}=t\slash\varepsilon \,\,\,\, {\rm and} \,\,\,\, r=r_{0}+\varepsilon \slash(h\,y),
\end{equation}
where 
\begin{equation}
h=(1+w)^{3}\slash(32M_{H}^{2}w^{2})
             \label{h_expr}
\end{equation} 
and $r_{0}=r_{extr}.$ 
Expanding the
function $f(r)$ in  powers  of $\varepsilon ,$ retaining quadratic terms and
subsequently taking the limit $\varepsilon \,=\,0$ we obtain 
\begin{equation}
ds^{2}\,=\,\frac{1}{hy^{2}}\left( -dt^{2}\,+\,dy^{2}\right)
\,+\,r_{0}^{2}d\Omega ^{2}.  \label{brdeformed2}
\end{equation}
Since $h^{-1}\,>\,r_{0}^{2},$ the line element does not belong to the
Bertotti-Robinson class, contrary to the near-horizon geometry of the 
Reissner-Nordstr\"{o}m solution.  Alternatively, this can easily be 
demonstrated making use of the relation
\begin{equation}
f''(r_{+}) = \frac{2}{r_{+}^{2}}+8\pi T_{\mu}^{\mu},
\end{equation}
where prime denotes differentiation with respect to the radial coordinate,
as the stress-energy tensor of the nonlinear electromagnetic field, 
$T_{\mu}^{\nu},$ has nonvanishing trace at the event horizon. 

Other frequently used representations of the line element (\ref{brdeformed2}) 
can be obtained through the change of coordinate system. Using, for example, 
\begin{equation}
h^{1/2}t\,=\,e^{\tau }\coth \chi ,\hspace{5mm}h^{1/2}y\,=\,e^{\tau }\sinh
^{-1}\chi  \label{set1}
\end{equation}
and 
\begin{equation}
\sinh ^{2}\chi \,=\,\mathcal{R}h-1,\hspace{5mm}\tau h\,=\,\mathcal{T}
\label{set2}
\end{equation}
one obtains
\begin{equation}
ds^{2}\,=\,\frac{1}{h}\left( -\sinh ^{2}\chi dt^{2}\,+\,d\chi ^{2}\right)
\,+\,r_{0}^{2}d\Omega ^{2}  \label{br1}
\end{equation}
and 
\begin{equation}
ds^{2}\,=\,-\left( \mathcal{R}^{2}h\,-\,1\right) d\mathcal{T}^{2}\,+\,\frac{d
\mathcal{R}^{2}}{\mathcal{R}^{2}h\,-\,1}\,+\,r_{0}^{2}d\Omega ^{2},
\label{br2}
\end{equation}
respectively. 
Topologically the geometry described by the line element (\ref{brdeformed2}) is
a direct product of the two-dimensional anti-de Sitter geometry and
the two-sphere of constant curvature; its curvature scalar
is simply a sum of the curvatures of the subspaces $
\mathrm{AdS}_{2}$ and $\mathrm{S}^{2}:$ 
\begin{equation}
R\,=\,K_{\mathrm{AdS}_{2}}\,+\,K_{\mathrm{S}^{2}},
\end{equation}
where $K_{\mathrm{AdS}_{2}}\,=\,-2h$ and $K_{\mathrm{S}^{2}}\,=
\,2/r_{0}^{2}. $

Now, let us return to the stress-energy tensor of the massive fields. Simple
calculations yield 
\begin{equation}
\langle T_{\mu }^{\nu }\rangle _{ren}^{\left( s\right) }=\frac{1}{96\pi
^{2}m^{2}}\,\mathrm{diag}\left[ A^{\left( s\right) },A^{\left( s\right)
},B^{\left( s\right) },B^{\left( s\right) }\right] _{\mu }^{\nu },
                  \label{eeextr}
\end{equation}
where 
\begin{equation}
A^{\left( 1/2\right) }=\frac{1}{21}h^{3}+\frac{1}{60r_{0}^{2}}h^{2}+\frac{1}{
42r_{0}^{6}},  \label{extr1}
\end{equation}
\begin{equation}
B^{\left( 1/2\right) }=-\frac{1}{42}h^{3}-\frac{1}{60r_{0}^{4}}h-\frac{1}{
21r_{0}^{6}}  \label{extr1a}
\end{equation}
and 
\begin{equation}
A^{\left( 1\right) }=\frac{8}{35}h^{3}+\frac{1}{5r_{0}^{2}}h^{2}+\frac{4}{
35r_{0}^{6}},
\end{equation}
\begin{equation}
B^{\left( 1\right) }=-\frac{4}{35}h^{3}-\frac{1}{5r_{0}^{4}}h-\frac{8}{
35r_{0}^{6}}  \label{extr4}
\end{equation}
for spinor and vector fields, respectively. In view of our 
earlier discussion we expect that the results (\ref{eeextr}-\ref{extr4})
coincide with the components of the stress-energy tensor
calculated at the event horizon of the extremal 
ABGB black hole. 

\subsection{Stress-energy tensor of massive spinor and vector fields in the
spacetime Reissner-Nordstr\"{o}m black hole}

The renormalized stress-energy tensor of the quantized massive 
spinor and vector fields
in the Reissner-Nordstr\"{o}m spacetime has been constructed in Ref. \cite
{Jirinek00prd}. 
It turns out that although the general formulas describing
$\langle T_{\mu }^{\nu }\rangle _{ren}^{\left( s\right) }$
are rather complicated, its components 
calculated in the Reissner-Nordstr\"{o}m spacetime 
are simple functions of the radial coordinate due to 
the spherical symmetry and the form of the metric potentials. 
These results will be used for the comparison with the 
analogous results obtained in the ABGB spacetime and 
we reproduce them here for the reader's convenience.

The components of the spinor field read 
\begin{eqnarray}
\langle T_{t}^{t}\rangle _{RN}^{(1/2)}\, &=&\,\frac{1}{40320\,\pi
^{2}\,m^{2}\,x^{12}M_{H}^{6}}\,\left( 2384\,{x}^{3}+10544\,{x}^{2}{q}
^{4}-22464\,{x}^{3}{q}^{2}+21832\,{x}^{2}{q}^{2}\right.   \notag \\
&-&\left. \,1080\,{x}^{4}-21496\,x{q}^{4}+4917\,{q}^{6}+5400\,{x}^{4}{q}
^{2}\right) ,  \label{t12}
\end{eqnarray}
\begin{eqnarray}
\langle T_{r}^{r}\rangle _{RN}^{(1/2)}\, &=&\,\frac{1}{40320\pi
^{2}\,m^{2}\,x^{12}M_{H}^{6}}\,\left( 504\,{x}^{4}+1080\,{x}^{4}{q}^{2}-784\,
{x}^{3}-6336\,{x}^{3}{q}^{2}\right.  \notag \\
&+&\left. \,3560\,{x}^{2}{q}^{4}+8440\,{x}^{2}{q}^{2}-8680\,x{q}^{4}+2253\,{q
}^{6}\right) ,  \label{r12}
\end{eqnarray}
and 
\begin{eqnarray}
\langle T_{\theta }^{\theta }\rangle _{RN}^{(1/2)}\, &=&\,-\frac{1}{
40320\,\pi ^{2}\,m^{2}\,x^{12}M_{H}^{6}}\left( -3536\,{x}^{3}+12080\,{x}^{2}{
q}^{4}-20016\,{x}^{3}{q}^{2}+30808\,{x}^{2}{q}^{2}\right.  \notag \\
&+&\left. \,1512\,{x}^{4}-33984\,xq^{4}+9933\,{q}^{6}+3240\,{x}^{4}{q}
^{2}\right) .  \label{a12}
\end{eqnarray}
Similarly, for the massive vector field one has 
\begin{eqnarray}
\langle T_{t}^{t}\rangle _{RN}^{(1)}\, &=&\,\frac{1}{10080\,\pi
^{2}\,m^{2}\,x^{12}M_{H}^{6}}\,\left( 31057\,{q}^{6}+1665\,{x}^{4}+41854\,{x}
^{2}{q}^{4}+93537\,{x}^{2}{q}^{2}\right.  \notag \\
&-&\left. \,107516\,x{q}^{4}-3666\,{x}^{3}-69024\,{x}^{3}{q}^{2}+12150\,{q}
^{2}{x}^{4}\right) ,  \label{t1}
\end{eqnarray}
\begin{eqnarray}
\langle T_{r}^{r}\rangle _{RN}^{(1)}\, &=&\,\frac{1}{10080\,\pi
^{2}\,m^{2}\,x^{12}M_{H}^{6}}\,\left( 1050\,{x}^{3}-693\,{x}^{4}+12907\,{x}
^{2}{q}^{2}-10448\,{x}^{3}{q}^{2}\right.   \notag \\
&-&\left. \,16996\,x{q}^{4}+2430\,{q}^{2}{x}^{4}+6442\,{x}^{2}{q}^{4}+5365\,{
q}^{6}\right) ,  \label{r1}
\end{eqnarray}
and 
\begin{eqnarray}
\langle T_{\theta }^{\theta }\rangle _{RN}^{(1)}\, &=&\,-\,\frac{1}{
10080\,\pi ^{2}\,m^{2}\,x^{12}M_{H}^{6}}\,\left( 13979\,{q}^{6}-2079\,{x}
^{4}+20908\,{x}^{2}{q}^{4}+30881\,{x}^{2}{q}^{2}\right.   \notag \\
&-&\left. \,44068\,x{q}^{4}+4854\,{x}^{3}-31708\,{x}^{3}{q}^{2}+7290\,{q}^{2}
{x}^{4}\right) .  \label{a1}
\end{eqnarray}
Although there are no numeric calculations of the stress-energy 
tensor of the quantized  massive
spinor and vector fields against which one could test the results (\ref{t12}-
\ref{a1}), we expect that the approximation is reasonable so long the mass
of the field is sufficiently large. Thanks to the detailed analytical 
and numerical calculations carried out in Refs.~\cite{Anderson1,Anderson2}
we know that this is indeed the case for the massive scalar field.
It is a very important result, indicating that that the exact stress-energy 
tensor of the scalar field may satisfactorily be approximated 
with the accuracy within a few percent provided $M_{H} m \geq 2$.
Further, as the sixth-order WKB-approach employed
in~\cite{Anderson1,Anderson2} is equivalent to the Schwinger-DeWitt 
expansion in inverse powers of $m^{2},$ this affirmative result yields 
a positive verification of the latter approach.

Finally, let us consider  the stress-energy tensor of the massive
fields in the spacetime of the extreme Reissner-Nordstr\"om black hole.
Its horizon value is given by
\begin{equation}
\langle T_{\mu}^{\nu}\rangle_{ren}^{(s)}\,=\,
\frac{\beta^{(s)}}{3360 \pi^{2} m^{2} M_{H}^{6}}
{\rm diag}[1,1,-1,-1],
              \label{ber_rob}
\end{equation} 
where $\beta^{(1/2)} = 37/12$ and $\beta^{(1)} =19.$
It can be easily demonstrated that it coincides with 
the stress-energy tensor 
of the massive field in the Bertotti-Robinson geometry.

\subsection{Massive spinor fields in ABGB spacetime}
                  \label{sec_spin}

Now, let us return to ABGB geometry and consider 
$\langle T_{\mu }^{\nu }\rangle _{ren}^{\left( s\right)
}$ near the event horizon of the extremal black hole. It can be shown that
the renormalized stress-energy tensor of the massive spinor field for $x$
close to $x_{extr}$ may be approximated by 
\begin{equation}
\langle T_{\mu }^{\nu }\rangle _{ren}^{\left( 1/2\right) }=\frac{\left(
1+w\right) ^{6}}{315\times 8^{7}\pi ^{2}m^{2}M_{H}^{6}w^{6}}\left[ A_{\mu
}^{\left( 1/2\right) \nu }+\frac{1}{2w}B_{\mu }^{\left( 1/2\right) \nu
}\left( x-x_{extr}\right) \right] +O\left( x-x_{extr}\right) ^{2},
\label{x12}
\end{equation}
where 
\begin{equation}
A_{t}^{\left( 1/2\right) t}=A_{r}^{\left( 1/2\right)
r}=57+44w+37w^{2}+10w^{3},  \label{A12}
\end{equation}
\begin{equation}
A_{\theta }^{\left( 1/2\right) \theta }=A_{\phi }^{\left( 1/2\right) \phi
}=-(99+29w+15w^{2}+5w^{3})  \label{AA12}
\end{equation}
and 
\begin{equation}
B_{t}^{\left( 1/2\right) t}=B_{r}^{\left( 1/2\right) r}=-\left( w+1\right)
(w+3)(52+7w+15w^{2}),  \label{B21}
\end{equation}
\begin{equation}
B_{\theta }^{\left( 1/2\right) \theta }=B_{\phi }^{\left( 1/2\right) \phi
}=\left( w+1\right) \left( 355+156w+58w^{2}+17w^{3}\right) .  \label{BB12}
\end{equation}
Numerically, one has 
\begin{eqnarray}
\langle T_{\mu }^{\nu }\rangle _{ren}^{\left( 1/2\right) } &=&\frac{1}{%
m^{2}M_{H}^{6}}10^{-4}\mathrm{diag}\left[ 1.039,\,1.039,\,-1.556,\,-1.556%
\right] _{\mu }^{\nu }  \notag \\
&&-\frac{1}{m^{2}M_{H}^{6}}10^{-4}\mathrm{diag}\left[ 5.958,\,5.958,\,0.780,
\,0.780\right] _{\mu }^{\nu }\left( x-x_{extr}\right) +O\left(
x-x_{extr}\right) ^{2}.  \label{num12}
\end{eqnarray}
To this end, observe that $x \to x_{extr}$ limit of $\ $(\ref{x12})
coincides with the stress-energy tensor of the massive spinor field in $
\mathrm{AdS}_{2}\times \mathrm{S}^{2}$ spacetime. To demonstrate this, it 
suffices to substitute into Eqs. (\ref{extr1}) and (\ref{extr1a}) the  explicit forms
of $r_{0}$ and $h$ as given by Eqs (\ref{xextr}) and (\ref{h_expr}), respectively.

Having established the expansion of the components of the stress-energy tensor
for the extremal configuration let us analyze their leading behavior for $
q\ll1.$ It can be shown that expanding the stress-energy tensor in powers of $
q$ one obtains
\begin{equation}
\langle T_{\mu }^{\nu }\rangle _{ren}^{\left( 1/2\right) }=\langle T_{\mu
}^{\nu }\rangle _{RN}^{\left( 1/2\right) }+\frac{q^{6}}{\pi
^{2}m^{2}M_{H}^{6}}t_{\mu }^{\left( 1/2\right) \nu }+\mathcal{O}\left(
q^{8}\right)  \label{q12}
\end{equation}
where
\begin{equation}
t_{t}^{\left( 1/2\right) t}=-\frac{5133-4444x+945x^{2}}{3360x^{12}}{,}
\label{qA12}
\end{equation}
\begin{equation}
t_{r}^{\left( 1/2\right) r}=-\frac{250-189x+35x^{2}}{1120x^{12}}{,}
\end{equation}
and
\begin{equation}
t_{\theta }^{\left( 1/2\right) \theta }=t_{\phi }^{\left( 1/2\right) \phi }=
\frac{1775-1421x+280x^{2}}{2240x^{12}}{,}
\label{qC12}
\end{equation}
where $\langle T_{\mu }^{\nu }\rangle _{RN}^{\left( 1/2\right) }$ is given
by Eqs. (\ref{t12}-\ref{a12}). 
Inspection of (\ref{t12}-\ref{a12}) and (\ref{qA12}-\ref{qC12}) indicates 
that for $q\ll 1$ the stress-energy tensor of the
quantized spinor field constructed in the spacetime of the ABGB black hole is almost
indistinguishable form the analogous tensor evaluated in the Reissner-Nordstr\"om 
geometry as they differ by $\mathcal{O}\left(q^{6}\right)$ terms. 

Now, let us consider the leading behavior of the stress-energy tensor
at large distances ($r/r_{+}\gg 1$). After some algebra the expansion 
valid for any $q$ may be written as 
\begin{equation}
\langle T_{\mu }^{\nu }\rangle _{ren}^{\left( 1/2\right) }=\frac{1}{\pi
^{2}m^{2}M_{H}^{6}}\tilde{t}_{\mu }^{\left( 1/2\right) \nu }+\mathcal{O}
\left( x^{-11}\right)
\end{equation}
where
\begin{equation}
\tilde{t}_{t}^{\left( 1/2\right) t}={\frac{3\left( 5\,{q}^{2}-1\right) }{112{
x}^{8}}}+{\frac{149-1404\,{q}^{2}}{2520{x}^{9}}}+\,{\frac{{q}^{2}\left(
2636\,{q}^{2}+5458-2835\,{q}^{4}\right) }{10080{x}^{10}},}  \label{s12rt}
\end{equation}
\begin{equation}
\tilde{t}_{r}^{\left( 1/2\right) r}=\,{\frac{15\,{q}^{2}+7}{560\mathit{x}^{6}
}}-\,{\frac{396\,{q}^{2}+49}{2520{x}^{9}}}+\,{\frac{{q}^{2}\left( 178\,{q}
^{2}+422-63\,{q}^{4}\right) }{2016{x}^{10}{\pi }^{2}},}
\end{equation}
and
\begin{equation}
\tilde{t}_{\theta }^{\left( 1/2\right) \theta }=\tilde{t}_{\phi }^{\left(
1/2\right) \phi }=-\,{\frac{3\left( 15\,{q}^{2}+7\right) }{560{x}^{8}}}+\,{%
\frac{221+1251\,{q}^{2}}{2520{x}^{9}}}-{\frac{{q}^{2}\left( 3851+1510\,{q}%
^{2}-630\,{q}^{4}\right) }{5040{x}^{10}}.}  \label{s12ra}
\end{equation}
Once again, the leading behavior of $\langle T_{\mu
}^{\nu }\rangle _{ren}^{\left( 1/2\right) }$ as $r\rightarrow \infty, $
(which is governed by the first term in the above equations and strongly 
depends on $q$) is identical to the analogous behavior in 
the Reissner-Nordstr\"om case.  
On the other hand, substituting $q =2 w^{1/2}$ into Eqs.~(\ref{s12rt}-\ref{s12ra})
one obtains the expansion of the stress-energy tensor at large distances
from the event horizon of the extreme black holes. It should be noted, however,
that any comparison of the extremal  ABGB and  Reissner-Nordstr\"om black holes
should be interpreted with care as the extremality limit occurs for 
 different values of $q.$

\subsection{Massive vector fields in ABGB spacetime}
                  \label{sec_vect}

The calculations of the renormalized stress-energy tensor of the quantized
massive vector fields proceed along the same lines as for the spinor case.
Repeating the steps necessary to calculate the SET of the massive spinor
field and focusing attention on the narrow strip near the degenerate event
horizon of the extremal black hole, one has 
\begin{equation}
\langle T_{\mu }^{\nu }\rangle _{ren}^{\left( 1\right) }=\frac{\left(
1+w\right) ^{6}}{210\times 8^{6}\pi ^{2}m^{2}M_{H}^{6}w^{6}}\left[ A_{\mu
}^{\left( 1\right) \nu }+\frac{1}{2w}B_{\mu }^{\left( 1\right) \nu }\left(
x-x_{extr}\right) \right] +\mathcal{O}\left( x-x_{extr}\right) ^{2},
\label{expansion_vector}
\end{equation}
where 
\begin{equation}
A_{t}^{\left( 1\right) t}=A_{r}^{\left( 1\right) r}=27+26w+19w^{2}+4w^{3},
\end{equation}
\begin{equation}
A_{\theta }^{\left( 1\right) \theta }=A_{\phi }^{\left( 1\right) \phi
}=-2(24+10w+3w^{2}+w^{3})
\end{equation}
and 
\begin{equation}
B_{t}^{\left( 1/2\right) t}=B_{r}^{\left( 1/2\right) r}=-\left( w+1\right)
(w+3)(25+7w+6w^{2}),
\end{equation}
\begin{equation}
B_{\theta }^{\left( 1/2\right) \theta }=B_{\phi }^{\left( 1/2\right) \phi }=
\frac{1}{6}\left( w+1\right) \left( 1475+1318w+485w^{2}+66w^{3}\right) .
\end{equation}
Since the location of the event horizon as well as the value of $q_{extr}$
depend on the particular value of the Lambert function one can easily
determine numerical value of the components of the stress-energy
tensor on the event horizon. Making use of (\ref{expansion_vector}) one
obtains 
\begin{eqnarray}
\langle T_{\mu }^{\nu }\rangle _{ren}^{\left( 1\right) } &=&\frac{1}{
m^{2}M_{H}^{6}}10^{-4}\mathrm{diag}\left[ 6.171,\,6.171,\,-9.321,\,-9.321
\right] _{\mu }^{\nu }  \notag \\
&&-\frac{1}{m^{2}M_{H}^{6}}10^{-4}\mathrm{diag}\left[ 35.563,\,35.563,
\,63.376,\,63.376\right] _{\mu }^{\nu }\left( x-x_{extr}\right) +\mathcal{O}\left(
x-x_{extr}\right) ^{2}.
\end{eqnarray}
Using, once again, Eqs.(\ref{xextr}) and (\ref{h_expr}) one can 
easily demonstrate that the horizon
value of the stress-energy tensor (\ref{expansion_vector}) 
reduces to that calculated in 
$\mathrm{AdS}_{2}\times \mathrm{S}^{2},$ geometry.

For any value of the radial coordinate and  small $q,$  the stress-energy 
tensor may be approximated by
\begin{equation}
\langle T_{\mu }^{\nu }\rangle _{ren}^{\left( 1\right) }=\langle T_{\mu
}^{\nu }\rangle _{RN}^{\left( 1\right) }+\frac{q^{6}}{\pi ^{2}m^{2}M_{H}^{6}}
t_{a}^{\left( 1\right) b}+\mathcal{O}\left( q^{8}\right)
\end{equation}
where
\begin{equation}
t_{t}^{\left( 1\right) t}=-\frac{(212249-172752x+34020x^{2})}{13440x^{12}}{,}
\end{equation}
\begin{equation}
t_{r}^{\left( 1\right) r}=-\frac{25859-19208x+3780x^{2}}{13440x^{12}}
\end{equation}
and
\begin{equation}
t_{\theta }^{\left( 1\right) \theta }=t_{\phi }^{\left( 1\right) \phi }=
\frac{82501-71316x+15120x^{2}}{13440x^{12}}.
\end{equation}
Now, expanding the general stress-energy tensor of the vector
field for $r/r_{+}\gg 1$ one obtains the leading terms (valid for any $q$) 
in the form
\begin{equation}
\langle T_{\mu }^{\nu }\rangle _{ren}^{\left( 1\right) }=\frac{1}{\pi
^{2}m^{2}M_{H}^{6}}\tilde{t}_{\mu }^{\left( 1\right) \nu }+\mathcal{O}\left(
x^{-11}\right) ,
                 \label{veve1}
\end{equation}
where 
\begin{equation}
\tilde{t}_{t}^{\left( 1\right) t}=\,{\frac{270\,{q}^{2}+37}{224{x}^{8}}}-{
\frac{11504\,{q}^{2}+611}{1680{x}^{9}}}-\,{\frac{{q}^{2}\left( -41854\,{q}
^{2}+25515\,{q}^{4}-93537\right) }{10080{x}^{10}},}
                       \label{vevet}
\end{equation}
\begin{equation}
\tilde{t}_{r}^{\left( 1\right) r}={-}\,{\frac{77-270\,{q}^{2}}{1120{x}^{8}}}-
{\frac{5224\,{q}^{2}-525}{5040{x}^{9}}}-\,{\frac{{q}^{2}\left( -12907-6442\,{
q}^{2}+2835\,{q}^{4}\right) }{10080{x}^{10}},}
\end{equation}
and
\begin{equation}
\tilde{t}_{\theta }^{\left( 1\right) \theta }=\tilde{t}_{\phi }^{\left(
1\right) \phi }=\,{\frac{3\left( 77-270\,{q}^{2}\right) }{1120{x}^{8}}}+{
\frac{15854\,{q}^{2}-2427}{5040{x}^{9}}}+\,{\frac{{q}^{2}\left( -20908\,{q}
^{2}+11340\,{q}^{4}-30881\right) }{10080{x}^{10}}.}
                      \label{veve4}
\end{equation}
Finally observe that the $q=2 w^{1/2}$ limit taken in Eqs. (\ref{veve1}-\ref{veve4}) 
supplements the discussion of the extremal black holes.

The results presented in Sec.~\ref{sec_spin} and~\ref{sec_vect} 
can be applied in further calculations. 
In the proofs of various theorems in General Relativity,
for example, the stress-energy tensor is expected to satisfy some 
restrictions usually addressed to as the energy conditions.
Their detailed studies are worthwhile as the violation
of the energy conditions frequently leads to exotic,
yet physically interesting situations. Of course, the main 
role played by the renormalized stress-energy tensor is to serve
as the source term of the semi-classical Einstein field equations. 
For the problem on hand one can calculate the back
reaction on the metric in the first-order approximation. 
Unfortunately, the components of the metric tensor of the quantum-corrected 
spacetime are rather complicated functions of the radial coordinate, 
each consisting of several hundred terms~\cite{unpubl}.
Therefore, to analyze the quantum-corrected spacetime it is necessary
to refer to approximations or even to numerical calculations.

\subsection{Inside the event horizon of the extremal ABGB black hole}

In this subsection we shall analyze the stress-energy tensor inside the
extremal ABGB black hole. The line element inside the degenerate horizon is
regular, and, for $r\rightarrow 0$\ it behaves as 
\begin{equation}
f\sim 1-\frac{4}{x}\exp \left( -4w/x\right) .  \label{asymp1}
\end{equation}
This may be contrasted with the analogous behavior of the
Reissner-Nordstr\"{o}m solution 
\begin{equation}
f\sim \frac{1}{x^{2}}.  \label{asymp1a}
\end{equation}

Even without detailed calculations certain qualitative features of $\langle
T_{\mu }^{\nu }\rangle _{ren}^{\left( s\right) }$ can be deduced from this
formulas. Indeed, since the stress-energy tensor is constructed form the
Riemann tensor, its covariant derivatives up to certain order and
contractions, the result of all this operations, in view of  the asymptotic
relation (\ref{asymp1}), should be regular. This can also be demonstrated
using Eq. (\ref{structure}), which, in the case in hand, can be written in
the form 
\begin{equation}
\langle T_{\mu }^{\nu }\rangle _{ren}^{\left( s\right) }=\frac{1}{96\pi
^{2}m^{2}M_{H}^{6}}\left( 1-\tanh \frac{2w}{x}\right) \sum_{i,j,k}\tilde{%
\alpha}_{ijk}^{\left( s\right) }\frac{w^{i}}{x^{j}}\tanh ^{k}\frac{2w}{x},
                      \label{asymp2a}
\end{equation}
where for each component $\tilde{\alpha}_{ijk}^{\left( s\right) }$ are
numerical coefficient depending on the spin of the massive field (we have
omitted tensor indices to make the formulas more transparent).
Alternatively, one can utilize approximation of the components of
the stress-energy tensor valid small $r$ 
\begin{equation}
\langle T_{\mu }^{\nu }\rangle _{ren}^{\left( s\right) }\sim \frac{1}{48\pi
^{2}m^{2}M_{H}^{6}}\exp \left( -4w/x\right) \sum_{i,j}\tilde{\beta}%
_{ij}^{\left( s\right) }\frac{w^{i}}{x^{j}}.  \label{asymp2}
\end{equation}
Inspection of Eqs.~(\ref{asymp2a}) or (\ref{asymp2}) shows that
$\langle T_{\mu }^{\nu }\rangle _{ren}^{\left( s\right) }\to 0$
as $r \to 0.$ This is simply because the Schwinger-DeWitt approximation
is local and depends on the geometric terms constructed from
the curvature. Since the line element has the Euclidean asymptotic
as $r \to 0 ,$ then, regardless of the spin of the field,
the renormalized stress-energy tensor must vanish in that limit.

It should be noted, however, that the regularity of the source term does not
necessarily leads to the regularity of the quantum corrected geometry. Indeed,
the latter requires that various curvature invariants of the self-consistent
solution of the semi-classical equations with the total source term given by the 
sum of classical stress-energy tensor of the nonlinear electrodynamics and 
of the quantized massive fields be regular. However, since the resulting 
semi-classical  equations comprise a very complicated system of sixth-order 
differential equations, there are no simple way to construct the appropriate solutions.
A comprehensive discussion of the analogous situation in the quadratic
gravity has been carried out  in~\cite{BeMaTry}.

\subsection{Numerical results}

Considerations of the previous sections concentrated on the approximate
analytical results valid in a few important regimes: $q\ll 1$, $x\gg 1$ and
for extremal configuration. Now, to gain insight into the overall behavior
of the stress-energy tensor as a function of $r$ and $q,$ one has to refer
to numerical calculations, as our complete but rather complicated results
are, unfortunately, not very illuminating. Below we describe the main
features of the constructed tensors and present them graphically. Related
discussion of the spin 0 field has been carried out in Refs.\cite
{Jirinek01prd1,Jirinek02prd}. 

First, let us consider the horizon values of
the components of $\langle T_{\mu }^{\nu }\rangle _{ren}^{\left( s\right) }.$ 
Spherical symmetry and regularity impose severe constrains on the
structure of the stress-energy tensor at the event horizon. It suffices,
therefore, to consider only its two independent components, say, $\langle
T_{t}^{t}\rangle _{ren}^{\left( s\right) }$ and $\langle T_{\theta }^{\theta
}\rangle _{ren}^{\left( s\right) }.$ The run of this components as functions
of $q$ is exhibited in Figs. 1 and 2, for spinor and vector fields,
respectively.

The run of the stress-energy tensor for a several exemplary
values of $q$ is exhibited in Figs. 3-13. Each curve represents
the radial dependence of the rescaled component of 
$\langle T_{\mu }^{\nu }\rangle _{ren}^{\left( s\right) } $ for a given $q.$ 
We shall start our discussion of the numerical results with the spin-$1/2$ field.
First, observe that
the energy density $\rho ^{\left( 1/2\right) }\left( \rho ^{\left( s\right)
}=-\langle T_{t}^{t}\rangle _{ren}^{\left( s\right) }\right) $ is always
negative at the event horizon, and, thus, by continuity, it is negative in its
vicinity. Further $\rho ^{\left( 1/2\right) }$ attains a positive local
maximum as can be clearly seen in Fig 3. For $q>1/\sqrt{5}$ the energy
density develops a negative minium (Fig. 4) and goes to $0^{-\text{ }}$as $
r\rightarrow \infty $. Further, inspection of the leading behavior of Eq.~(\ref{s12rt})
shows that $\rho ^{\left( 1/2\right) }$ is
positive at large distances  for $q<1/\sqrt{5.}$ 

The radial pressure $p_{r}^{\left( 1/2\right) } 
$ $(p_{r}^{\left( s\right) }=\langle T_{r}^{r}\rangle _{ren}^{\left(
s\right) })$ is positive at the event horizon and $p_{r}^{\left( 1/2\right)
}\left( r_{+}\right) =-\rho ^{\left( 1/2\right) }$ $\left( r_{+}\right) $;
subsequently, it decreases monotonically to $0^{+}$ with $r$. The behavior of 
$p_{r}^{\left( 1/2\right) }$ is plotted in Fig. 5. 

The tangential pressure, $p_{\theta }^{\left( 1/2\right) }
\left( p_{\theta }^{\left( s\right)
}=\langle T_{\theta }^{\theta }\rangle _{ren}^{\left( s\right) }\right) $ 
is plotted in Figs. 6 and 7.
At the event horizon it is positive for $q<0.823$, approaches a negative minimum
at $r/r_{+}\approx 1.5$ and goes to $0^{-\text{ }}$ as $r\rightarrow \infty.$ 
A closer examination indicates that for $q> 0.756,$ it develops a local maximum,
which disappears near the extremality limit. 

In general, there are no  qualitative similarities between 
components of the renormalized stress-enegy tensor of the massive spinor
and vector fields, as can be easily seen in Figs. 8-13.
In the vicinity of the event horizon the energy density of the massive
vector field is positive for $q<0.581$ and negative otherwise.  For $
q>0.465$ the energy density approaches a  maximum, and, subsequently, 
regardless of $q$ it has a minimum. As the leading behavior as $
r\rightarrow \infty $ is governed by the first term in rhs of (\ref{vevet}), $
p_{t}^{\left( 1\right) }\rightarrow 0^{-}.$ Other qualitative and
quantitative features of the energy density can easily be inferred from
Figs. 8 and 9. 

Numerical calculations indicate that for $q<0.387$ the radial pressure, $p_{r}^{\left(
1\right) },$ is negative and monotonically increases to $0^{-}$ as $r$ $
\rightarrow \infty .$  For $0.387<q<$ $0.919$ there
appears a local minimum in the closest neighborhood of the event horizon,
and, for $0.534<q<0.919$ , the radial pressure approaches a local maximum.
Finally, for $q>0.919$, it decreases monotonically to $0^{+}.$ The run of $
p_{r}^{\left( 1\right) }$ for a few exemplary values of $q$ is plotted in
Figs. 10 and 11. 

The tangential pressure of the vector field is negative on
the event horizon and increases to a global maximum. Subsequent behavior of $
p_{\theta }^{\left( 1\right) }$ depends on $q$: it decreases monotonically
to $0^{+\text{ }}$ for $q<0.534$ whereas for $q>0.534$ the tangential
pressure has a local minimum and increases to $0^{-}.$ Some other
qualitative and quantitative features, as for example the numerical values
of $p_{\theta }^{\left( 1\right) }$ at the maxima and minima can easily be
inferred from Figs. 12 and 13.

The numerical calculations carried out in the external region of the extremal
configuration  shows that the run of the stress-energy tensor
qualitatively  follows the analogous behavior for $q=1$ case, and,
consequently, it will not be discussed separately. 

Now, let us consider the vacuum polarization effects inside the event
horizon of the extremal configuration. The run of the rescaled
components of stress-energy tensor of the massive spinor field
is exhibited in Figs 14-16. All the components display oscillatory behavior
for $r/r_{+}>0.05,$ indicating that the back reaction effects would 
be especially interesting there.
Such a behavior can easily be understood in relation with the behavior
of the line element. Indeed, for small $r$ the line element closely
resembles that of a flat spacetime, and, consequently, the vacuum
polarization effects are small.
On the other hand, for $r/r_{+}>0.05$ the function $f(r)$ changes noticeably
leading to the  changes of the stress-energy tensor.
The competition of the local geometric terms $\delta W_{i}/\delta g_{\mu\nu}$
lead to its oscillatory-like behavior.
Numerical calculations indicate that the stress-energy tensor of the quantized
vector field is qualitatively similar to that of the spinor field and 
approximately one has 
\begin{equation}
\langle T_{\mu}^{\nu}\rangle _{ren}^{(1) }\approx 10 \times 
\langle T_{\mu}^{\nu}\rangle _{ren}^{(1/2)} 
\end{equation}
The basic features of the stress-energy tensor of the quantized 
massive vector field can easily be infrred form Figs. 14-16 and the above relation.

\section{Final remarks}

In this paper we have constructed the renormalized stress-energy tensor of 
the massive spinor and vector fields in the spacetime of ABGB black 
hole. The scalar case has been analyzed extensively in our two previous 
papers. The method employed here is based on the observation that the first-order
effective action could be expressed in terms of the (traced) coincidence limit
of the coefficient $a_{3}.$ Functional differentiation of this action
with respect to the metric tensor yields the most general first-order 
(i.e. proportional to $m^{-2}$) stress-energy tensor. Such a generic 
tensors of the quantized massive scalar, spinor and vector fields 
have been constructed for the first time in~\cite{Jirinek00prd,Jirinek01prd1}. 

Application of our general formulas, although conceptually straightforward,
is technically rather intricate, and produces quite complex results.
Therefore, for clarity, we have analyzed the leading behavior of 
$\langle T_{\mu }^{\nu }\rangle _{ren}^{\left( s\right) }$
in some physically important regimes. This discussion has
been supplemented with detailed numerical calculations. 
The results have also been used to construct and analyze the stress-energy 
tensor in $\mathrm{AdS}_{2}\times \mathrm{S}^{2},$ spaces, 
which are naturally related to the near horizon geometry
of the extremal ABGB black hole.
 
A special emphasis in this paper has been put
on the extremal configurations.  Specifically, it has been shown that 
the stress-energy tensor of the massive fields is regular inside the degenerate
event horizon. This result raises important question of the nature of the
black hole interior  in the back-reaction problem. Preliminary calculations
carried out in~\cite{BeMaTry} for the quadratic gravity, 
which, for certain calculational purposes, may be considered as some sort
of a toy model of the semi-classical theory, indicate that at least for 
the first-order calculations it is possible to obtain regular solution, at the
expense of a small modification of the classical nonlinear action.
Of course, the stress-energy 
tensor of the quantized massive fields 
constructed in the general static, spherically symmetric
and asymptotically flat spacetime is far more complicated than quadratic 
terms~\cite{BeMaTry}, however, the general pattern that lies behind 
the calculations should be, in general, the same. The calculations
carried out so far indicate that this is indeed the case, although lengthy and
complicated results expressed in term of the polylogarithms are rather
hard to analyze and manipulate. 
Moreover, it would be interesting to investigate the back reaction 
problem for any $q$ outside the event horizon.
Finally, observe that the ABGB solution with the cosmological constant 
may provide an interesting setting for studying the influence of the
quantized fields upon ultraextremal horizons.
These problems are being studied and the results will be 
reported elsewhere.

\begin{figure}[h]
\includegraphics[width=8cm]{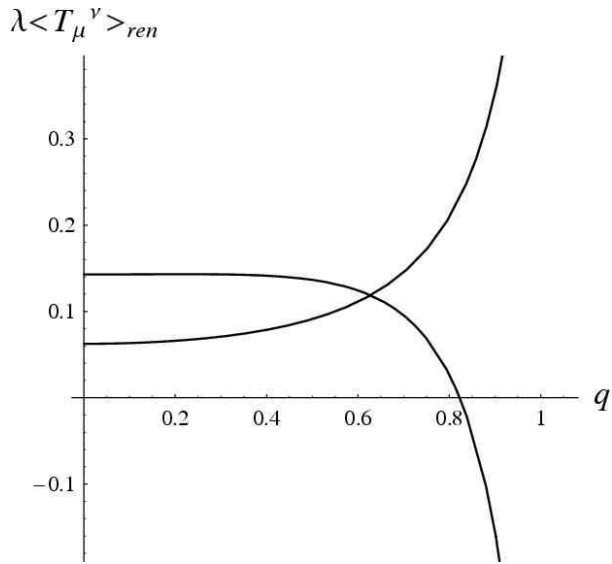}
\caption[sp_hor.ep]{This graph shows behavior of the rescaled components of $
\langle T_{t}^{t}\rangle_{ren}^{(1/2)}$ and $\langle
T_{\theta}^{\theta}\rangle_{ren}^{(1/2)}$ [$\protect\lambda =5760 \protect\pi^{2}
m^{2} M_{H}^{6}$] of the renormalized stress- energy tensor of the quantized
massive spinor field at the event horizon. The time component is always
positive and increases with $q,$ whereas the angular component is positive 
for $q<0.823.$ 
For the extremal configuration $
\lambda \langle T_{t}^{t}\rangle_{ren}^{(1/2)} = 5.907$ and $
\lambda \langle T_{\theta}^{\theta}\rangle_{ren}^{(1/2)} = -8.843.$
}
\label{fig1}
\end{figure}
\begin{figure}[h]
\includegraphics[width=8cm]{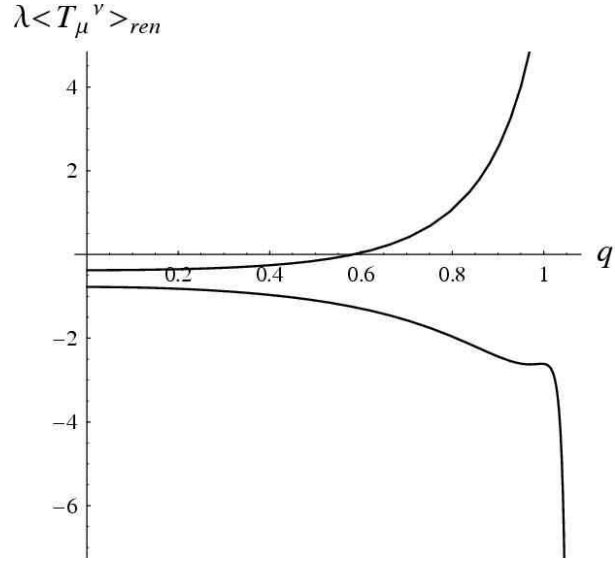}
\caption[vec_hor.ep]{This graph shows behavior of the rescaled components of 
$\langle T_{t}^{t}\rangle_{ren}^{(1)}$ and $\langle
T_{\theta}^{\theta}\rangle_{ren}^{(1)}$ [$\protect\lambda =5760 \protect\pi^{2} m^{2}
M_{H}^{6}$]of the renormalized stress- energy tensor of the quantized
massive vector field at the event horizon. The time component
increases with $q$ and is negative for $q< 0.581. $ For the extremal configuration
$\lambda \langle T_{t}^{t}\rangle_{ren}^{(1)} = 35.080$ and $
\lambda \langle T_{\theta}^{\theta}\rangle_{ren}^{(1)} = -52.990$} 
\label{fig2}
\end{figure}
\begin{figure}[h]
\includegraphics[width=8cm]{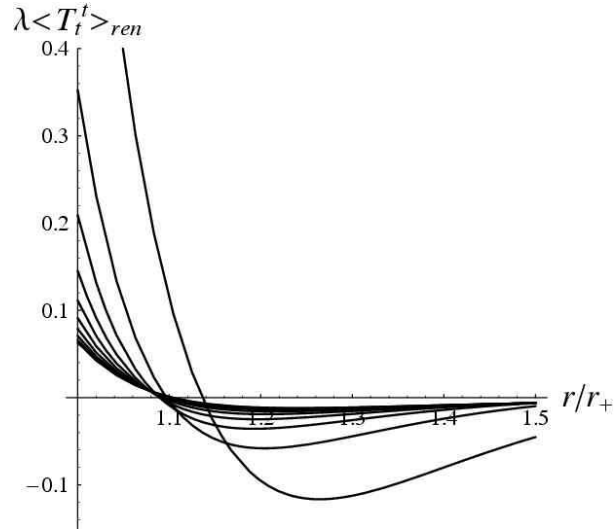}
\caption[sp_tt_a.ep]{This graph shows the radial dependence of the rescaled
component $\langle T_{t}^{t}\rangle_{ren}^{(1/2)}$ [$\protect\lambda =5760 
\protect\pi^{2} m^{2} M_{H}^{6}$] of the stress-energy tensor of the
quantized massive spinor field in the spacetime of the ABGB black hole. From
top to bottom at the event horizon the curves are plotted for $q = 1-i/10,$ $%
(i=0,1,...,9).$ Each curve attains a negative minimum in the vicinity of $%
r_{+}.$}
\label{fig3}
\end{figure}
\begin{figure}[h]
\includegraphics[width=8cm]{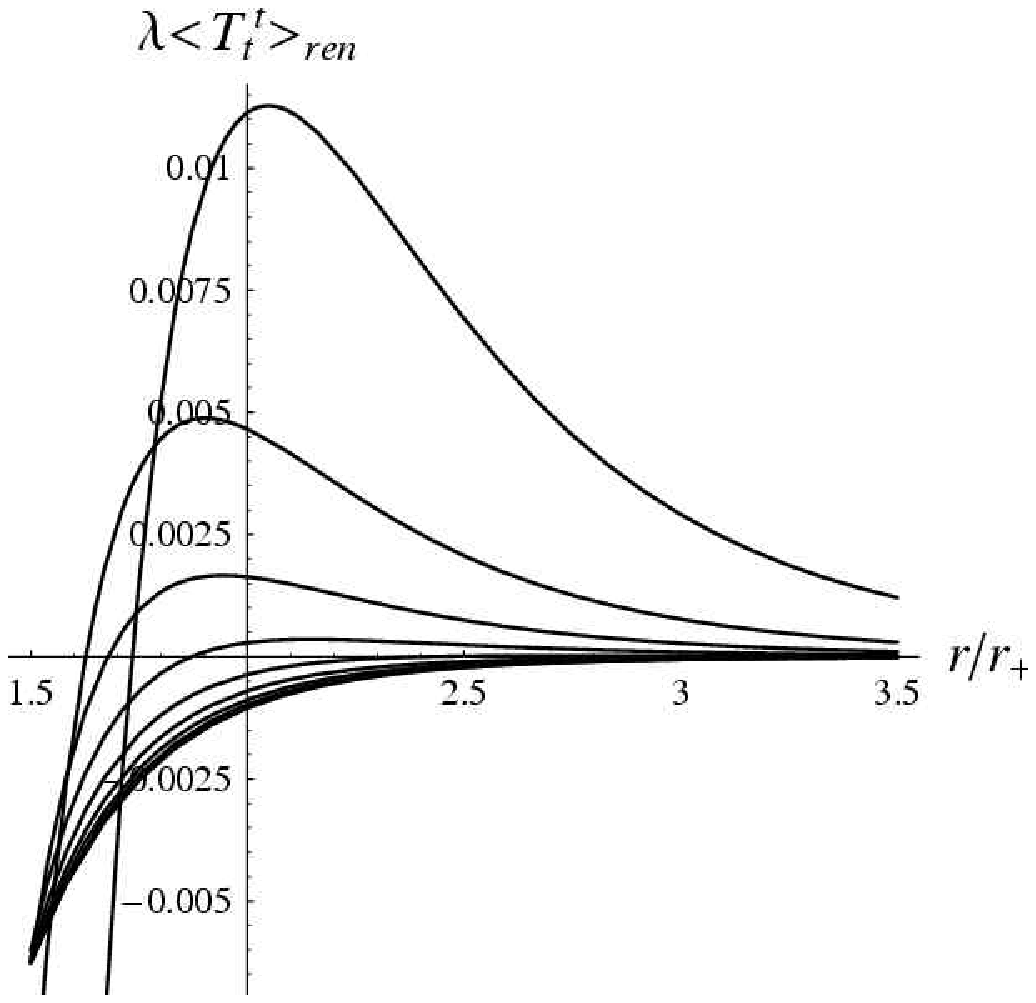} .
\caption[sp_tt_b.ep]{This graph shows the radial dependence of the rescaled
component $\langle T_{t}^{t}\rangle_{ren}^{(1/2)}$ [$\protect\lambda =5760 
\protect\pi^{2} m^{2} M_{H}^{6}$] of the stress-energy tensor of the
quantized massive spinor field in the spacetime of the ABGB black hole 
for $1.5<r/r_{+}<3.5$. From top to bottom the curves are plotted for 
$q = 1-i/10,$ $(i=0,...,9)$}
\label{fig4}
\end{figure}
\begin{figure}[h!]
\includegraphics[width=8cm]{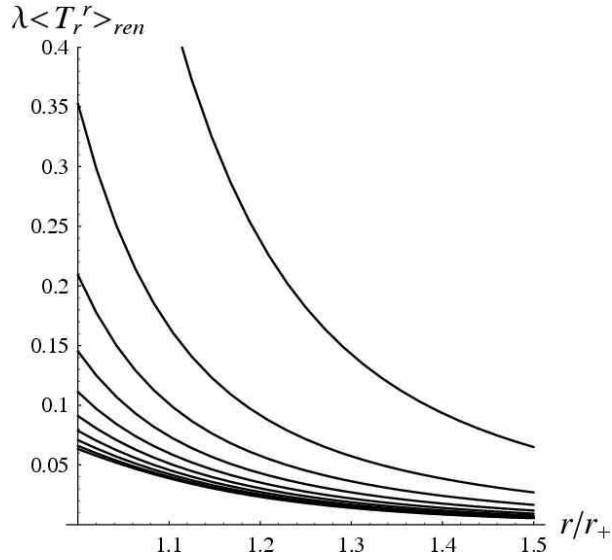}
\caption[sp_rr_a.ep]{ This graph shows the radial dependence of the rescaled
component $\langle T_{r}^{r}\rangle_{ren}^{(1/2)}$ [$\protect\lambda =5760 
\protect\pi^{2} m^{2} M_{H}^{6}$] of the stress-energy tensor of the
quantized massive spinor field in the spacetime of the ABGB black hole. From
top to bottom the curves are plotted for $q = 1-i/10,$ $(i=0,...,9).$ Each
curve decreases monotonically to $0^{+}$ with $r.$ }
\label{fig5}
\end{figure}
\begin{figure}[h!]
\includegraphics[width=8cm]{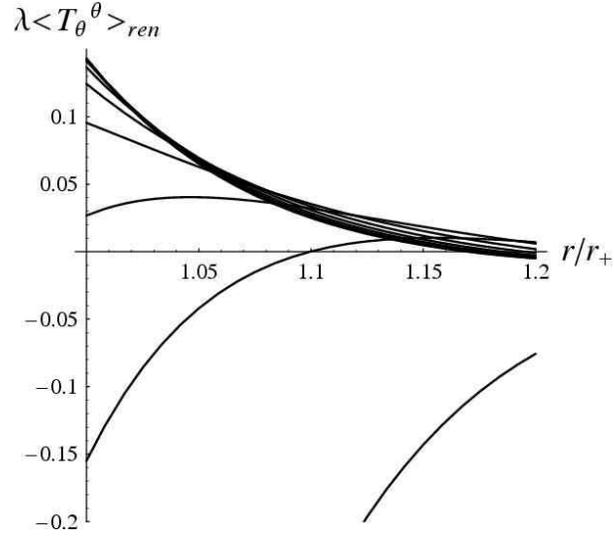}
\caption[sp_ang_a.ep]{ This graph shows the radial dependence of the
rescaled component $\langle T_{\protect\theta}^{\protect\theta%
}\rangle_{ren}^{(1/2)}$ [$\protect\lambda =5760 \protect\pi^{2} m^{2}
M_{H}^{6}$] of the stress-energy tensor of the quantized massive spinor
field in the spacetime of the ABGB black hole. From top to bottom at the
event horizon the curves are plotted for $q = i/10,$ $(i=1,...,10).$ For $q
< 0.823$ the component $\langle T_{\protect\theta}^{\protect\theta}\rangle_{ren}^{(1/2)}$
is positive in the vicinity of the event horizon. }
\label{fig6}
\end{figure}
\begin{figure}[h!]
\includegraphics[width=8cm]{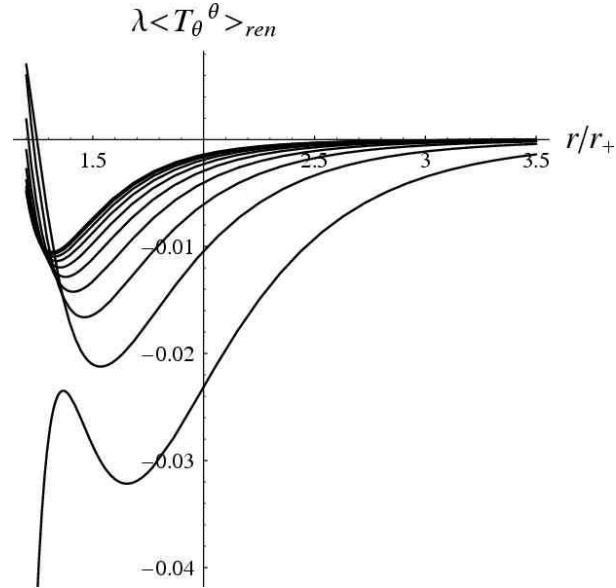}
\caption[sp_ang_b.ep]{This graph shows the radial dependence of the rescaled
component $\langle T_{\protect\theta}^{\protect\theta}\rangle_{ren}^{(1/2)}$
[$\protect\lambda =5760 \protect\pi^{2} m^{2} M_{H}^{6}$] of the
stress-energy tensor of the quantized massive spinor field in the spacetime
of the ABGB black hole for $1.2 < r/r_{+} <3.5.$ 
From top to bottom (in the minima) the functions are
plotted for $q = i/10,$ $(i=1,...,10).$}
\label{fig7}
\end{figure}
\begin{figure}[h!]
\includegraphics[width=8cm]{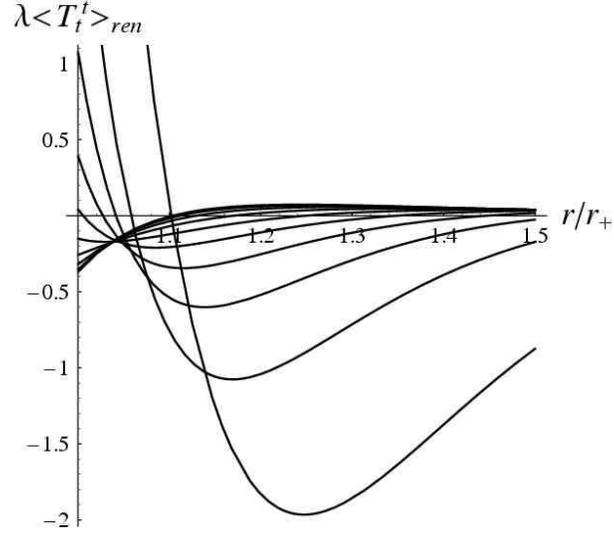}
\caption[vec_tt_a.ep]{This graph shows the radial dependence of the rescaled
component $\langle T_{t}^{t}\rangle_{ren}^{(1)}$ [$\protect\lambda =5760 
\protect\pi^{2} m^{2} M_{H}^{6}$] of the stress-energy tensor of the
quantized massive vector field in the spacetime of the ABGB black hole. At
the event horizon $\langle T_{t}^{t}\rangle_{ren}^{(1)}$ is positive for $%
q>0.851.$ For $q < 0.465$ the curves reach minimum.}
\label{fig8}
\end{figure}
\begin{figure}[h!]
\includegraphics[width=8cm]{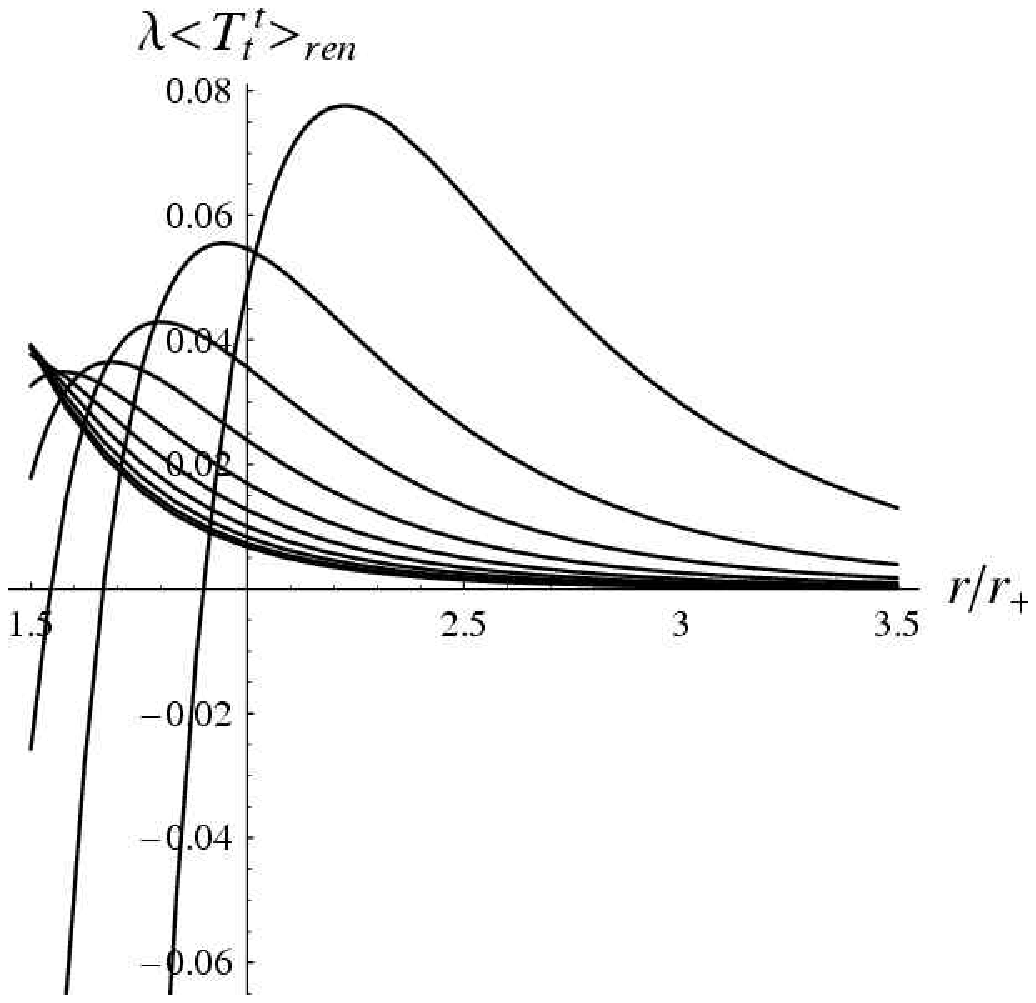}
\caption[vec_tt_b.ep]{This graph shows the radial dependence of the rescaled
component $\langle T_{t}^{t}\rangle_{ren}^{(1)}$ [$\protect\lambda =5760 
\protect\pi^{2} m^{2} M_{H}^{6}$] of the stress-energy tensor of the
quantized massive vector field in the spacetime of the ABGB black hole. From
top to bottom (at $r=2.5 r_{+}$) horizon the curves are plotted for $q =
1-i/10,$ $(i=0,...,9).$}
\label{fig9}
\end{figure}
\begin{figure}[h!]
\includegraphics[width=8cm]{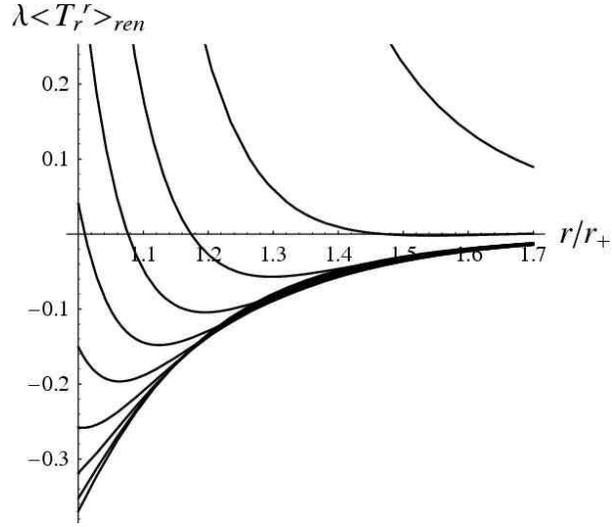}
\caption[vec_rr_a.ep]{ This graph shows the radial dependence of the
rescaled component $\langle T_{r}^{r}\rangle_{ren}^{(1)}$ [$\protect\lambda %
=5760 \protect\pi^{2} m^{2} M_{H}^{6}$] of the stress-energy tensor of the
quantized massive vector field in the spacetime of the ABGB black hole. From
top to bottom at the event horizon the curves are plotted for $q = 1-i/10,$ $%
(i=0,...,9).$ For $q < 0.387$ $\langle T_{r}^{r}\rangle_{ren}^{(1)}$
increases with $r$ to $0^{-}$ whereas for $q>0.919$ it is a monotonically
decreasing function.}
\label{fig10}
\end{figure}
\begin{figure}[h!]
\includegraphics[width=8cm]{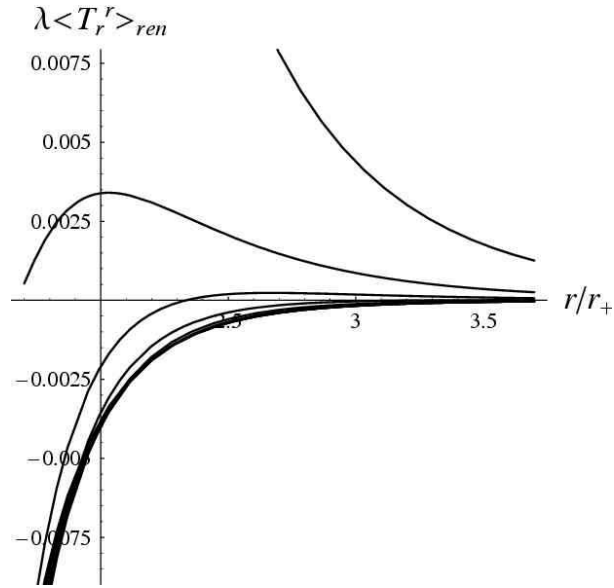}
\caption[vec_rr_b.ep]{ This graph shows the radial dependence of the
rescaled component $\langle T_{r}^{r}\rangle_{ren}^{(1)}$ [$\protect\lambda %
=5760 \protect\pi^{2} m^{2} M_{H}^{6}$] of the stress-energy tensor of the
quantized massive vector field in the spacetime of the ABGB black hole for $%
1.7 <r/r_{+}<3.7$}
\label{fig11}
\end{figure}
\begin{figure}[h!]
\includegraphics[width=8cm]{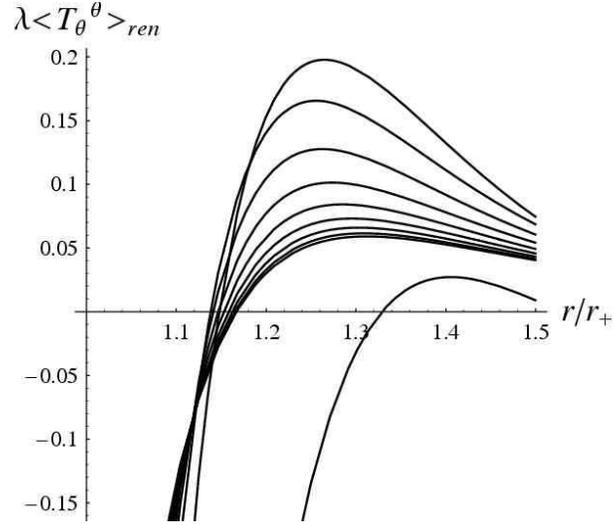}
\caption[vec_ang_a.ep]{ This graph shows the radial dependence of the
rescaled component $\langle T_{\protect\theta}^{\protect\theta%
}\rangle_{ren}^{(1)}$ [$\protect\lambda =5760 \protect\pi^{2} m^{2}
M_{H}^{6} $] of the stress-energy tensor of the quantized massive vector
field in the spacetime of the ABGB black hole. It is always negative at the
event horizon and increases to a local maximum. From top to bottom the
curves are for $q = 1-i/10,$ $(i=1,...,9)$ and $q=1.$}
\label{fig12}
\end{figure}
\begin{figure}[h!]
\includegraphics[width=8cm]{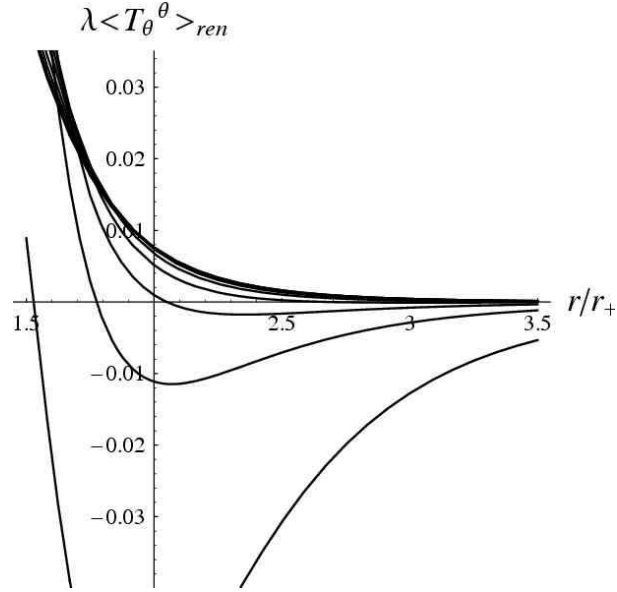}
\caption[vec_ang_b.ep]{ This graph shows the radial dependence of the
rescaled component $\langle T_{\protect\theta}^{\protect\theta%
}\rangle_{ren}^{(1)}$ [$\protect\lambda =5760 \protect\pi^{2} m^{2}
M_{H}^{6} $] of the stress-energy tensor of the quantized massive vector
field in the spacetime of the ABGB black hole. For $q<0.534$ it decreases
monotonically to $0^{+};$ for $q>0.534$ it has a local minimum and increases
to $0^{-}.$ }
\label{fig13}
\end{figure}
\begin{figure}[h!]
\includegraphics[width=8cm]{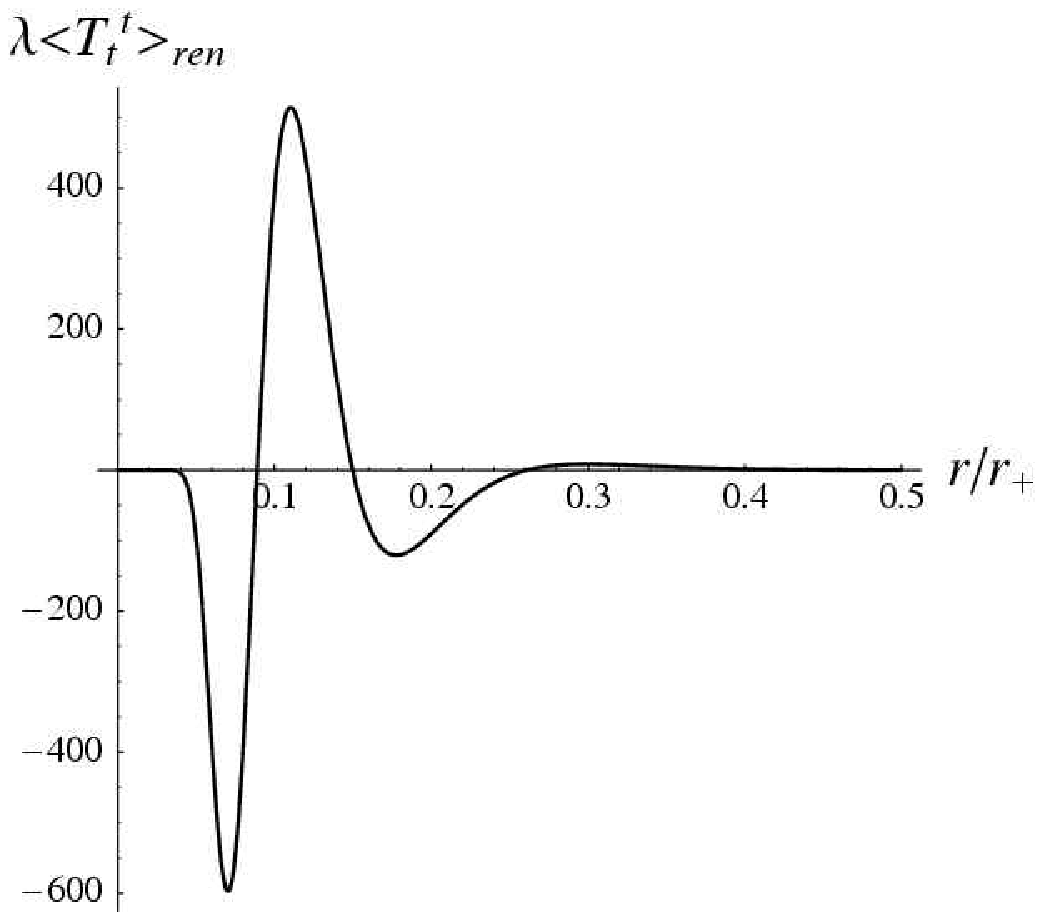}
\caption[vec_ang_b.ep]{ This graph shows the radial dependence of the
rescaled component $\langle T_{t}^{t} \rangle_{ren}^{(1)}$ [$\protect\lambda = m^{2}
M_{H}^{6} $] of the stress-energy tensor of the quantized massive spinor
field inside the event horizon of the extreme ABGB black hole.}
\label{fig14}
\end{figure}

\begin{figure}[h!]
\includegraphics[width=8cm]{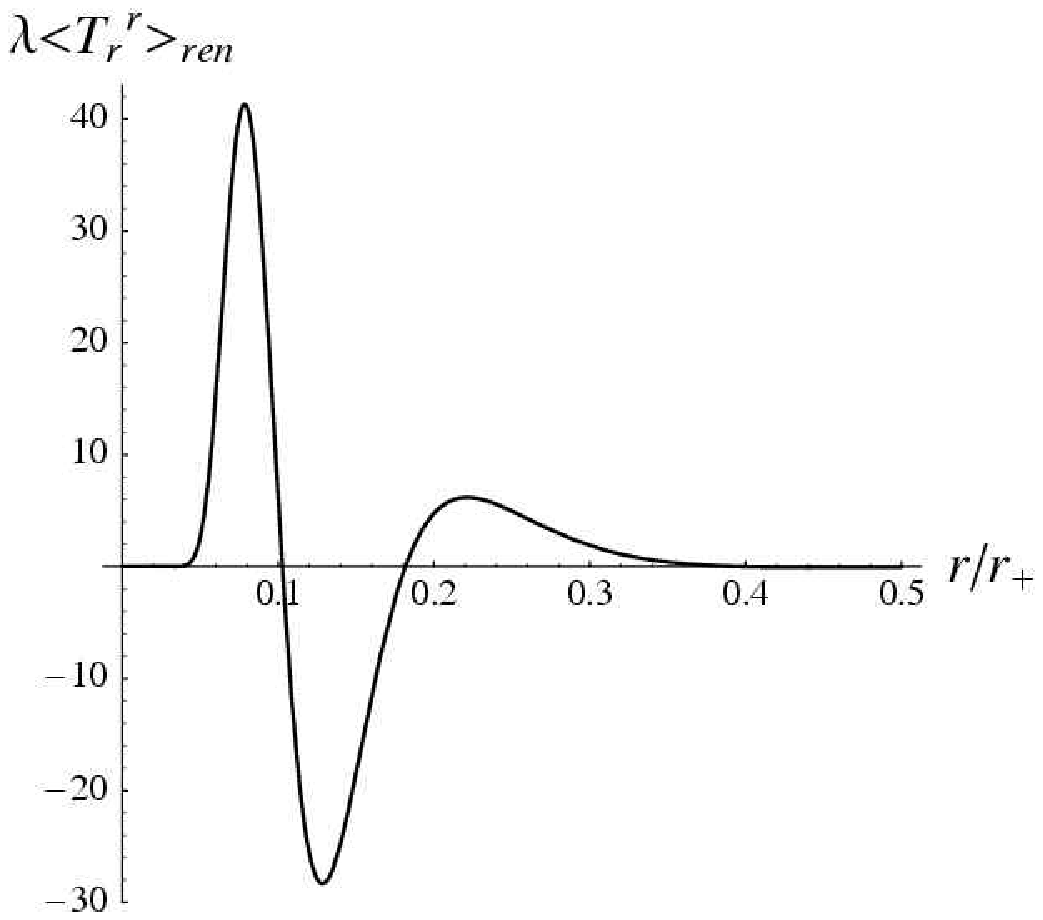}
\caption[vec_ang_b.ep]{ This graph shows the radial dependence of the
rescaled component $\langle T_{r}^{r}\rangle_{ren}^{(1)}$ [$\protect\lambda = m^{2}
M_{H}^{6} $] of the stress-energy tensor of the quantized massive spinor
field inside the event horizon of the extreme ABGB black hole.}
\label{fig15}
\end{figure}

\begin{figure}[h!]
\includegraphics[width=8cm]{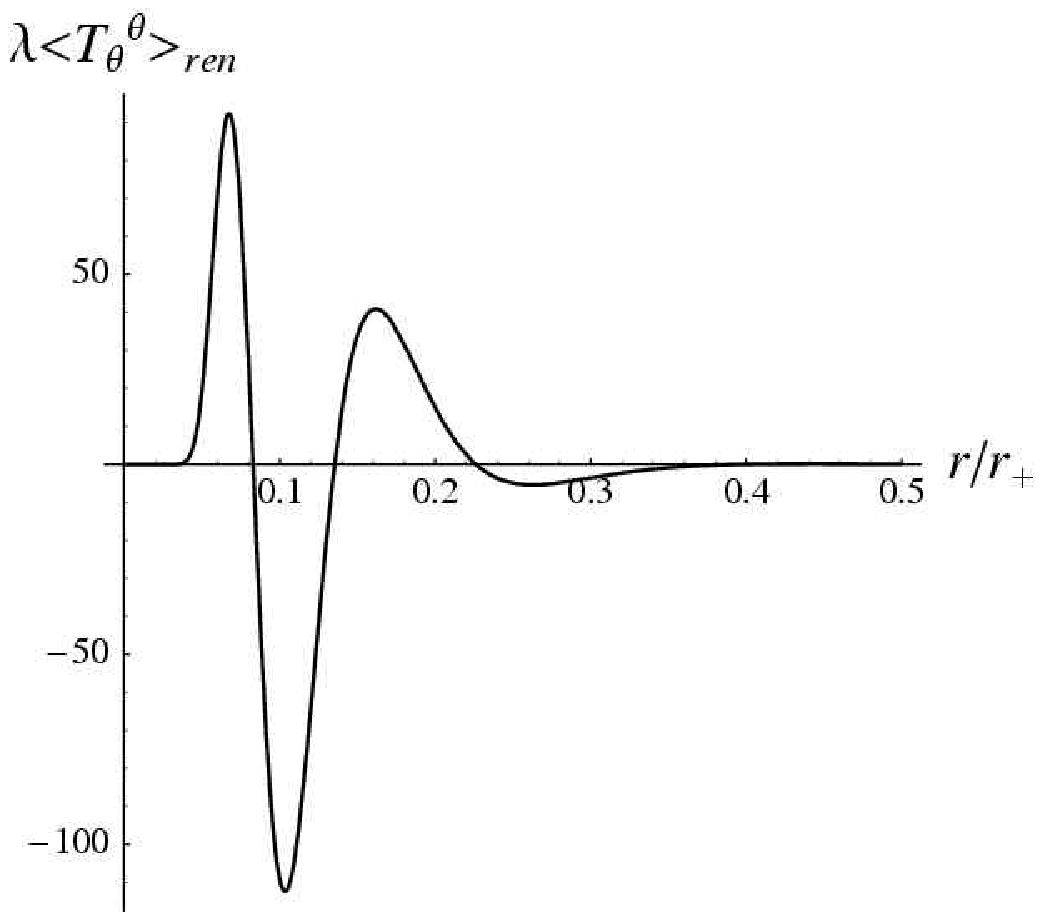}
\caption[vec_ang_b.ep]{ This graph shows the radial dependence of the
rescaled component $\langle T_{\protect\theta}^{\protect\theta%
}\rangle_{ren}^{(1)}$ [$\protect\lambda = m^{2}
M_{H}^{6} $] of the stress-energy tensor of the quantized massive spinor
field inside the event horizon of the extreme ABGB black hole.}
\label{fig16}
\end{figure}

\clearpage


\end{document}